\documentclass[aoas,MSNbibl,nameyear,seceqn,dvips]{arximspdf}
\usepackage{graphicx}
\doi{10.1214/14-AOAS779} 
\volume{8}
\issue{4}
\pubyear{2014}
\firstpage{2404}
\lastpage{2434}
\docsubty{FLA}

\makeatletter
\newcommand{\eqref}[1]{(\ref{#1})}
\newcommand{\bG}{\bolds{\Gamma}}
\newcommand{\bg}{\bolds{\gamma}}

\newcommand{\bD}{\bolds{\Delta}}

\newcommand{\bZ}{\mathbf {Z}}

\newcommand{\bmm}{\mathbf{ m}}
\newcommand{\bu}{\mathbf{ u}}

\makeatother

\begin{document}
\begin{frontmatter}

\title{Detecting duplicates in a homicide registry using a~Bayesian
partitioning approach\thanksref{T1}}
\runtitle{Bayesian Duplicate Detection}
\thankstext{T1}{Supported by the Carnegie Mellon
University node of the NSF Census Research Network (NCRN)
under NSF Grant SES-1130706.}

\begin{aug}
\author[A]{\fnms{Mauricio} \snm{Sadinle}\corref{}\ead[label=e1]{msadinle@stat.cmu.edu}}
\runauthor{M. Sadinle}

\affiliation{Carnegie Mellon University}

\address[A]{Department of Statistics\\
Carnegie Mellon University\\
Pittsburgh, Pennsylvania 15213\\
USA\\
\printead{e1}}
\end{aug}
%

\received{\smonth{4} \syear{2014}}
\revised{\smonth{7} \syear{2014}}

%
\begin{abstract}
Finding duplicates in homicide registries is an important step in
keeping an accurate account of lethal violence. This task is not
trivial when unique identifiers of the individuals are not available,
and it is especially challenging when records are subject to errors and
missing values. Traditional approaches to duplicate detection output
independent decisions on the coreference status of each pair of
records, which often leads to nontransitive decisions that have to be
reconciled in some ad-hoc fashion. The task of finding duplicate
records in a data file can be alternatively posed as partitioning the
data file into groups of coreferent records. We present an approach
that targets this partition of the file as the parameter of interest,
thereby ensuring transitive decisions. Our Bayesian implementation
allows us to incorporate prior information on the reliability of the
fields in the data file, which is especially useful when no training
data are available, and it also provides a proper account of the
uncertainty in the duplicate detection decisions. We present a study to
detect killings that were reported multiple times to the United Nations
Truth Commission for El Salvador.
\end{abstract}

%
\end{frontmatter}

\section{Introduction}\label{sec1}

Duplicate detection is the task of finding sets of records that refer
to the same entities within a data file. This task is not trivial when
unique identifiers of the entities are not recorded in the file, and it
is especially difficult when the records are subject to errors and
missing values. The existence of duplicates in a data file may
compromise the validity of any analysis that uses those data, and
therefore duplicate detection is needed in a wide variety of contexts,
including public health and biomedical research [e.g., \citet{Hsuetal00,Milleretal00,Sariyaretal12MissingValues}], and census
quality improvement [e.g., \citet{Fay04,Marshall08}].

In the context of an armed conflict, it is common for an institution
recording civilian casualties to receive multiple reports on the same
victims. These reports may come from witnesses who provide different
degrees of detail, therefore leading to nontrivial duplicates in the
institution's data file. Finding duplicates in those homicide
registries is an important step toward keeping an accurate account of
lethal violence. In this article we study a case from El Salvador,
where a Truth Commission formed by the United Nations in 1992 collected
data on killings that occurred during the Salvadoran civil war
(1980--1991). Due to the way in which those data were collected, a
victim could have been reported by different relatives and friends, and
therefore it is important to detect those multiply reported casualties.

\subsection{The United Nations Truth Commission for El Salvador}

From 1980 to 1991, the Republic of El Salvador, in Central America,
underwent a civil war between the Salvadoran Government and the
left-wing guerrilla Farabundo Mart\'i National Liberation Front (FMLN,
after its name in Spanish). The parties signed a peace agreement in
1992 which later led to the creation of the Commission on the Truth for
El Salvador by the United Nations [\citeauthor{Buergenthal94}
(\citeyear{Buergenthal94,Buergenthal96})],
henceforth abbreviated as UNTC.

Between 1992 and 1993, the UNTC summoned the Salvadoran society to
report violations that occurred during the war, mainly focusing on
homicides and disappearances of noncombatants. The UNTC ran
announcements on the radio, television, and in newspapers inviting
individuals to testify, and opened offices in different regions of the
country where information from witnesses was collected [\citet{CTElSalvador93}]. Finally, in 1993 the UNTC published a report with
the results of their investigations, including a list of homicides
directly obtained from testimonials, which were mainly provided by the
victims' family members, but also by close friends. In addition to the
names of the victims, this list contains the reported locations and
dates of the killings.

Most of the killings reported to the UNTC occurred several years before
1992, and therefore it was expected that friends and relatives of the
victims would not recall some details of the killings or would provide
testimonials that conflict with each other. These characteristics of
the data collection naturally led to missing information and nontrivial
duplicate records in the UNTC data file. The variability among records
that refer to the same victim and the presence of missing data make
finding duplicates specially challenging. Furthermore, it is difficult
to construct a reliable training data set for duplicate detection, that
is, a set of record pairs with known coreference statuses, which
supervised duplicate detection methods require. In this document we
develop a new approach to duplicate detection inspired by these type of
situations. Our approach handles missing data and allows the duplicate
detection process to be assisted with prior information on the
reliability of each field in the file, which helps to compensate for
the absence of training data.

\subsection{Current approaches to duplicate detection}

Duplicate detection differs from the closely related task of \emph
{record linkage} in the sense that the goal of the latter is to link
multiple files usually obtained from different data collection
processes, and it is assumed that these files do not contain duplicates
within them [\citet{FellegiSunter69,Winkler88,Jaro89,LarsenRubin01,HerzogScheurenWinkler07}]. Despite this difference, the same principles
and techniques can usually be adapted to solve both tasks.

In this article two or more records referring to the same entity are
called \textit{coreferent}. Traditional approaches to unsupervised
duplicate detection and record linkage fit mixture models on pairwise
comparisons of records with the goal of separating coreferent from
noncoreferent pairs [\citet{Elmagarmidetal07,HerzogScheurenWinkler07}].
Traditional supervised approaches train classifiers on a sample of
record pairs with known coreference statuses, and then predict the
coreference statuses of the remaining record pairs [\citet{Elmagarmidetal07,ChristenBook}]. Both of these type of approaches
output independent decisions on the coreference status of each record
pair, and therefore neither of them guarantee transitivity of the
coreference decisions. For example, it is possible that records $i$ and
$j$ are declared as being coreferent, as well as records $j$ and $k$,
but records $i$ and $k$ may be declared as noncoreferent. If $i$, $j$
and $k$ truly correspond to the same entity, the nontransitivity could
occur due to measurement error and incomplete record information. It
may be the case, however, that only two or none of those records are
coreferent, but these methodologies do not offer any representation of
uncertainty in these situations, and so they require resolving
discrepancies in an ad-hoc post-processing step.

Most recently, Bayesian approaches to both duplicate detection and
record linkage have been proposed, which provide a natural account of
the coreference decisions' uncertainty in the form of posterior
distributions. Most of these approaches directly model the information
contained in the data files [\citet{Matsakis10,TancrediLiseo11,Fortinietal02,Gutmanetal13,Steortsetal13}],
which require crafting specific models for each type of field in the
file, and are therefore currently limited to handle nominal categorical
fields or continuous variables modeled under normality. In practice,
however, fields that are complicated to model, such as names,
addresses, phone numbers or dates, are important to detect coreferent
records. These type of fields are often subject to typographical and
other types of errors, which make it important to take into account
partial agreements between their values. This is certainly an advantage
of traditional methodologies, as they base their decisions on pairwise
comparisons of records and therefore can use any type of field, as long
as these can be compared in a meaningful way. The approaches of
\citet{Fortinietal01} and \citeauthor{Larsen02}
(\citeyear{Larsen02,Larsen05,Larsen12}) are Bayesian
implementations of the traditional unsupervised approach to record
linkage [\citet{FellegiSunter69,Winkler88,LarsenRubin01,HerzogScheurenWinkler07}], which bases its coreference decisions on
pairwise comparison data. These latter approaches, however, do not
currently handle missing data, do not take into account multiple levels
of partial agreement, and they would lead to nontransitive decisions if
they were applied without modification to a duplicate detection problem.

\subsection{Overview of the article}

The approach that we propose in this article builds upon the previous
literature by combining a number of desirable characteristics for a
duplicate detection technique. Our approach to duplicate detection
guarantees transitivity of the coreference decisions by defining our
parameter of interest as the partition of the data file that groups
coreferent records together, as in \citet{Matsakis10}. Our approach is
closely related to those of \citet{FellegiSunter69}, \citet{Winkler88},
\citet{Jaro89}, \citet{LarsenRubin01}, \citet{Fortinietal01} and
\citeauthor{Larsen02} (\citeyear{Larsen02,Larsen05,Larsen12}) for record linkage in the sense that our
coreference decisions are based on comparison data, but we also extend
some ideas of \citet{Winkler90Strings} to take into account levels of
disagreement among the fields' values. In practice, it is also common
to have missing values in the data file, and so we show how our method
can be adapted to those situations. By taking a Bayesian approach we
can incorporate prior knowledge on the reliability of the fields, which
is useful in situations where no training data are available. The
introduction of prior information to solve this type of problem has
been advocated by \citet{Fortinietal01}, \citeauthor{Larsen02}
(\citeyear{Larsen02,Larsen05,Larsen12}) and others. Our Bayesian approach provides us with a
posterior distribution on the possible partitions of the file, which is
a natural way to account for the uncertainty in the coreference
decisions, similarly as in \citet{Matsakis10} and \citet{Steortsetal13}.

The remainder of the article is organized as follows: Section~\ref{s:methodology} presents a general description of the proposed
methodology; Section~\ref{s:modelindepcomps} presents a conditional
independence model that leads to a simple way of dealing with missing
values, an illustrative example and a simulation study; Section~\ref{s:UNTRCapplication} addresses the problem of detecting killings
reported multiple times to the United Nations Truth Commission for El
Salvador; and Section~\ref{s:conclusions} concludes.

\section{Methodology}\label{s:methodology}

Assume we have a data file containing $r$ records labeled $\{1,\ldots,r\}
$, where more than one record may refer to the same underlying entity.
Finding duplicates in such a data file is equivalent to grouping
records according to the underlying entities that they refer to. If
there are $n\leq r$ entities represented in the data file, we can
safely think of partitioning the data file into $n$ groups of
coreferent records. This partition of the file, called \textit{coreference
partition} [\citet{Matsakis10}], is our parameter of interest, and it
can be represented in different ways. We use different representations
throughout the article depending on which one is more convenient.

\subsection{Representations of partitions}\label{ss:representations}

A partition of a set is a collection of nonempty and nonoverlapping
subsets whose union is the original set. In this article those subsets
are called \textit{groups} or \textit{cells}. Given a data file with, say,
five records $\{1,2,3,4,5\}$, a partition with cells $\{1,3\}$, $\{2\}$
and $\{4,5\}$ is denoted as $1, 3 / 2 / 4, 5$. In a coreference
partition, each of its cells represents an underlying entity,
therefore, in this example records 1 and 3 are coreferent, as well as
records 4 and 5. This representation, however, is not useful for computations.

A partition can also be represented by a matrix. Let us consider the
matrix $\bD$ of size $r \times r$, whose $(i,j)$th entry is defined as
\[
\Delta_{ij}= \cases{ %
1, &\quad  $\mbox{if records
$i$ and $j$ refer to the same entity;}$
\vspace*{2pt}\cr
0, &\quad $\mbox{otherwise.}$}
\]
In the context of duplicate detection we will refer to $\bD$ as a \emph
{coreference matrix}. Notice that $\bD$ is symmetric with only ones in
the diagonal, and it would be block-diagonal if coreferent records were
contiguous in the data file, with each block representing a group of
coreferent records.

Representing partitions using matrices is computationally inefficient,
especially when the number of records is large. An alternative is to
use arbitrary labelings of the partition's cells. Since $r$ is the
number of records in the data file, it is safe to assume that $r$ is
the maximum number of entities possibly represented in the data file,
and therefore it is the maximum number of labels that we need. By
assigning an arbitrary labeling to these $r$ potential entities, we can
introduce the variables $Z_i$, $i=1,\ldots,r$, where $Z_{i}=q$ if record
$i$ represents entity $q$, with $1\leq q \leq r$, and the vector $\bZ
=(Z_1,\ldots,Z_r)$ contains all the records' labels. Notice that
although the labeling of the $r$ potentially existing entities is
arbitrary, any relabeling leads to the same partition of the records.
In fact, $\Delta_{ij}=I(Z_i=Z_j)$, where $I(\cdot)$ is the indicator
function, and this relationship does not depend on the labeling that we
use. This relationship is important since a prior distribution on the
space of partitions can be obtained by specifying a distribution for
the records' labels $\bZ$. Notice that if the number of entities $n$ is
lower than $r$, then there will be $r-n$ labels not in use for each
particular labeling. According to this labeling scheme, a partition of
$r$ elements into $n$ cells has $r!/(r-n)!$ possible labelings.
Finally, to fix ideas, the vectors $\bZ=(1, 2, 1, 3, 3)$ and $\bZ=(4,
1, 4, 2, 2)$ are instances of arbitrary labelings of the partition $1,
3 / 2 / 4, 5$, since in both $Z_1=Z_3\neq Z_4=Z_5$, and $Z_2$ gets its
own unique value.

The number of ways in which a data file with $r$ records can be
partitioned is given by the $r$th Bell number [see, e.g., \citet{Rota64}], which grows rapidly with $r$. For example, the number of
possible partitions of a file with 10 records is 115,975, and if the
file contains 15 records, the Bell number grows to 1,382,958,545. In
practice, most files are much larger, but fortunately most partitions
can be ruled out at an early stage, as we describe in Section~\ref{ss:complexity}. To make inferences on the file's coreference
partition, we find how similar each pair of records is.

\subsection{Levels of disagreement as comparison data}\label{ss:compdata}

Comparison data are obtained by comparing pairs of records, with the
goal of finding evidence of whether two records refer to the same
entity. Intuitively, two records referring to the same entity should be
very similar. The way of constructing the comparisons depends on the
information contained by the records. The most straightforward way of
comparing the same field of two records is by checking whether their
information agrees or not. Although this comparison method is
extensively used, and it is appropriate for comparing unordered
categorical fields (e.g., sex or race), it completely ignores partial agreement.

\citet{Winkler90Strings} proposes to take into account partial agreement
among fields that contain strings (e.g., given names) by computing a
string metric, such as the normalized Levenshtein edit distance or any
other [see \citet{Bilenkoetal03,Elmagarmidetal07}], and then dividing
the resulting set of similarity values into different \textit{levels of
disagreement}. Winkler's approach can be extended to compute levels of
disagreement for fields that are not appropriately compared in a
dichotomous fashion.

We compare the field $f$ of records $i$ and $j$ by computing some
similarity measure $\mathcal{S}_f(i,j)$. The range of this similarity
measure is then divided into $L_f+1$ intervals $I_{f0}, I_{f1},\ldots,
I_{fL_f}$, that represent different levels of disagreement. By
convention, the interval $I_{f0}$ represents the highest level of
agreement, which includes no disagreement, and the last interval,
$I_{fL_f}$, represents the highest level of disagreement, which
depending on the field represents complete or strong disagreement. We
can then build ordinal variables from these intervals. For records $i$
and $j$, and field $f$, we define
\[
\gamma^{f}_{ij} = l\qquad \mbox{if } \mathcal{S}_f
(i,j) \in I_{fl}.
\]
The larger the value of $\gamma^{f}_{ij}$, the larger the disagreement
between records $i$ and $j$ with respect to field $f$. These different
field comparisons are collected in a vector for each record pair, as in
the record linkage literature [e.g., \citet{FellegiSunter69}]. $\bg
_{ij}=(\gamma_{ij}^1,\ldots,\gamma_{ij}^f,\ldots,\gamma_{ij}^F)$ denotes
the comparison vector for records $i$ and $j$, where $F$ is the number
of fields being compared.

Notice that, in principle, we could construct $\bg_{ij}$ using the
original similarity values $\mathcal{S}_f(i,j)$. In our approach,
however, we model these comparison vectors as a way to make inference
on the coreference partition. Modeling directly the original $\mathcal
{S}_f(i,j)$'s requires a customized model per type of comparison, since
these similarity measures output values in different ranges, depending
on their functional form and the field being compared. By building
levels of disagreement as ordinal categorical variables, we can use a
generic model for any type of comparison, as long as its values are categorized.

This approach also raises the question of how to choose the thresholds
to build the intervals $I_{fl}$. The selection of the thresholds should
correspond to what the researcher genuinely considers as levels of
disagreement. This depends on the specific application at hand and the
type of field being compared. For example, in Sections~\ref{s:modelindepcomps} and \ref{s:UNTRCapplication} we build levels of
disagreement according to what we consider to be no disagreement, mild
disagreement, moderate disagreement and extreme disagreement.

Although in principle the number of record comparisons is $
{r\choose 2}=r(r-1)/2$, in practice, most record pairs are noncoreferent, and
most of them can be trivially detected using some simple criteria,
thereby avoiding the computation of the complete set of comparisons, as
we show next.

\subsection{Reducing the inferential and computational complexity}\label
{ss:complexity}

In most applications there are simple ways to detect large numbers of
obvious noncoreferent pairs at some early stage of the duplicate
detection process. Detecting those pairs reduces tremendously the
inferential and computational complexity of the problem, given that
whenever records $i$ and $j$ are declared as noncoreferent, this
translates to fixing $\Delta_{ij}=0$ in the coreference matrix, which
in turn assigns probability zero to all the partitions where records
$i$ and $j$ are grouped together.

There are different techniques to detect sets of noncoreferent pairs,
and here we refer to a few of them [see \citet{Christen12} for an
extensive survey]. The most popular approach is called \textit{blocking},
and it consists of dividing the data file into different blocks (sets
of records) according to one or more reliable categorical fields, such
that records in different blocks are considered to be noncoreferent.
The idea is that if a field is reliable enough, then it would be
unlikely to find a coreferent pair among pairs of records disagreeing
in that field. For example, if we believe a field like gender or postal
code (zip code) to be free of error, we can declare records disagreeing
on that field to be noncoreferent. This approach is appealing since it
does not even require us to compute comparisons, as the file can be
simply divided according to the categories of the fields being used for
blocking.

In many cases no field may be completely trusted, and therefore
blocking may lead to miss truly coreferent pairs. We can, however,
exploit prior knowledge on the types of errors expected for the
different fields. By understanding what kind of errors would be
unlikely for a certain field, we can declare as noncoreferent any pair
of records that disagrees by more than a predefined threshold with
respect to the field in consideration. Ideally, this comparison should
be cheap to compute, since it will be checked for all record pairs. For
example, information on time events for individuals, such as date of
birth or date of death, is misreported in certain contexts, but it is
common that whenever the correct date is not recorded, the date that
appears in the record is somehow close to the true one. In this
example, two records containing dates that are very different could be
declared as noncoreferent. Other fields that can be used in this
fashion include age or geographic information, given that in many
contexts it is unlikely to find coreferent pairs among records that
report very different ages or distant locations. Naturally, the
validity of any of these approaches has to be assessed on a
case-by-case basis.

Ideally after applying one of the previous steps, or a combination of
them, the set of pairs is reduced to a manageable size for which
complete comparisons can be computed, as explained in Section~\ref{ss:compdata}. Computationally expensive comparisons, such as those
involving string metrics, should be reserved for this stage. We call
$\mathcal{P}$ the set of pairs for which complete comparisons are
computed. The comparison data for the pairs in $\mathcal{P}$ comprises
the information that we will use to estimate the partition of the file.
Within $\mathcal{P}$, however, many pairs may still be obvious
noncoreferent pairs that can be detected using combinations of the
different levels of disagreement.\vadjust{\goodbreak} Therefore, we can further reduce the
complexity of the inferential task by declaring record pairs as
noncoreferent whenever they strongly disagree according to some
user-defined criteria built using the computed levels of disagreement.
For instance, criteria for declaring a pair as noncoreferent could be
having strong disagreements in given and family names, or having strong
disagreements in a combination of fields such as age, race and
occupation, if they were available. Finally, if a pair of records meet
any of the established criteria, then it is declared as noncoreferent.
The reasoning behind this approach is that, although no single field
may be enough to distinguish further noncoreferent records, strong
disagreements in a combination of fields are probably a good indication
of the records being noncoreferent, and therefore we would expect this
approach to be robust to errors. The set of remaining pairs whose
coreference statuses are still unknown is denoted by~$\mathcal{C}$, and
we refer to it as the set of candidate pairs. Although we fix the pairs
in $\mathcal{P}-\mathcal{C}$ as noncoreferent, we use their comparison
data in the model presented in the next section, since those pairs
provide examples of noncoreferent records.

The possible coreference partition of the file is now constrained to
the set $\mathcal{D}=\{\bD\dvtx \Delta_{ij}=0, \ \forall (i,j)\notin
\mathcal{C}\}$, that is, the set of partitions that do not group
together the record pairs that have already been declared as
noncoreferent. In practice, $\mathcal{D}$ is much smaller than the set
of all possible partitions of the file, which is why we heavily rely on
being able to have a small set of candidate pairs $\mathcal{C}$ to
apply our method to medium or large size data files.

\subsection{Model description}\label{ss:GeneralModel}

We now present a model for the comparison data $\bg= \{\bg_{ij}\}
_{(i,j)\in\mathcal{P}}$ such that the distribution of the comparison
vectors depends on whether the pairs are coreferent or not, which will
allow us to estimate the coreference partition. Notice that we model
all the pairs in $\mathcal{P}$ even though those in $\mathcal
{P}-\mathcal{C}$ are fixed as noncoreferent.

We assume that the comparison vector $\bg_{ij}$ is a realization of a
random vector $\bG_{ij}$, and the comparison data $\bg$ are a
realization of a random array $\bG$. It is clear that the set of record
pairs is composed of two types: coreferent and noncoreferent pairs.
Furthermore, we expect the distribution of the comparison vectors $\bG
_{ij}$ to be very different among those two types. For example, we
expect to observe more agreements among coreferent pairs than among
noncoreferent pairs and, similarly, we expect many more disagreements
among noncoreferent pairs than among coreferent pairs. This intuition
can be formalized by assuming that the distribution of $\bG_{ij}$ is
the same for all record pairs that refer to the \emph{same} entity
(regardless of the entity), and that the distribution of $\bG_{ij}$ is
the same for all record pairs that refer to \emph{different} entities
(regardless of the pair of entities). These assumptions have been
widely employed for linking different data files under the
Fellegi--Sunter framework for record linkage [\citet{FellegiSunter69,Winkler88,LarsenRubin01,HerzogScheurenWinkler07}].

The intuitive description above can be formalized into a model for the
comparison data as
%
\begin{eqnarray}
\label{eq:model1} \bG_{ij}|\Delta_{ij}&=&1\stackrel{\mathrm{i.i.d.}} {
\sim} G_1,
\nonumber
\\[-8pt]
\\[-8pt]
\nonumber
\bG_{ij}|\Delta_{ij}&=&0\stackrel{\mathrm{i.i.d.}} {\sim}
G_0,
\end{eqnarray}
for all $(i,j)\in\mathcal{P}$, where $G_1$ and $G_0$ represent the
models of the comparison vectors for pairs that are coreferent and
noncoreferent, respectively. These models have to be specified
according to the comparison data at hand. Leaving $G_1$ and $G_0$
unspecified by now, we can see that for a configuration of the
coreference matrix~$\bD$, the joint probability of observing the
comparison data $\bg$ can be written as
%
\begin{eqnarray}\label{eq:PG_D}
&&\mathbb{P}(\bG= \bg| \bD, \Phi)
\nonumber
\\[-8pt]
\\[-8pt]
\nonumber
&&\qquad= \prod_{(i,j)\in\mathcal{C}}
\mathbb {P}_1(\bg_{ij}|\Phi_1)^{\Delta_{ij}}
\mathbb{P}_0(\bg_{ij}|\Phi _0)^{1-\Delta_{ij}}
\prod_{(i,j)\in\mathcal{P}-\mathcal{C}} \mathbb {P}_0(
\bg_{ij}|\Phi_0),
\end{eqnarray}
where $\mathbb{P}_1(\bg_{ij}|\Phi_1):=\mathbb{P}(\bG_{ij}=\bg
_{ij}|\Delta_{ij}=1,\Phi_1)$ and, similarly, $\mathbb{P}_0(\bg_{ij}|\Phi
_0):=\mathbb{P}(\bG_{ij}=\bg_{ij}|\Delta_{ij}=0,\Phi_0)$, with $\Phi
=(\Phi_1,\Phi_0)$ representing a parameter vector of the models $G_1$
and $G_0$. Notice that equation \eqref{eq:PG_D} is obtained given that
we fix $\Delta_{ij}=0$ for those pairs in $\mathcal{P}-\mathcal{C}$.
Also, although the posterior on $\bD$ does not depend directly on the
comparison data for pairs in $\mathcal{P}-\mathcal{C}$, it does depend
on $\Phi_0$, which in turn depends on those pairs in $\mathcal
{P}-\mathcal{C}$. In fact, the previous formulation is equivalent to a
model for only the candidate pairs $\mathcal{C}$, as long as the factor
$\prod_{(i,j)\in\mathcal{P}-\mathcal{C}} \mathbb{P}_0(\bg_{ij}|\Phi_0)$
gets incorporated in the prior for $\Phi_0$.

\subsection{Prior distribution on the coreference partition}\label
{ss:PriorCorefMatrix}

Since the coreference matrix $\bD$ represents a partition, the entries
of $\bD$ are not independent, for example, if $\Delta_{ij}=1$ and
$\Delta_{jk}=1$, then $\Delta_{ik}=1$. In a mixture model
implementation of the model presented in equations \eqref{eq:model1}
and \eqref{eq:PG_D}, the $\Delta_{ij}$'s ($i<j$) are taken as i.i.d.
Bernoulli($p$), where $p$ represents the proportion of coreferent pairs
[\citet{Elmagarmidetal07,Sariyaretal09,SariyarBorg10,ChristenBook}]. The
independence assumption of the $\Delta_{ij}$'s in a mixture model
approach to duplicate detection leads to nontransitive decisions on the
coreference statuses of record pairs. To avoid these undesirable
results, we treat $\bD$ as a partition and put a prior distribution on
it accordingly.

As we showed before, $\mathcal{D}$ denotes the set of possible
coreference partitions. In this article we use the prior that assigns
equal probability to each partition in $\mathcal{D}$. This flat prior
is such that $\pi(\bD)\propto I(\bD\in\mathcal{D})$. We can also
obtain this prior in terms of the partition labelings introduced in
Section~\ref{ss:representations}. The set $\mathcal{D}$ is equivalent
to the set of labelings
$\mathcal{Z}=\{\bZ\dvtx  Z_i\neq Z_j,\ \forall  (i,j)\notin\mathcal{C}\}
$. A simple way to obtain the flat prior for $\bD$ from a prior for $\bZ
$ is by assigning equal probability to each of the $r!/(r-n)!$
labelings of a partition with $n$ cells, which leads to the prior on
labelings $\pi(\bZ)\propto[(r-n(\bZ))!/r!]I(\bZ\in\mathcal{Z})$,
where $n(\bZ)$ measures the number of different labels in labeling $\bZ$.

Notice that in some situations it may be desired to use a more
structured prior on partitions, for example, if the researcher has a
prior idea about the percentage of duplicates. How to appropriately
incorporate this information requires further investigation, since
commonly used distributions on partitions encourage the formation of
large cells as they are designed for traditional clustering problems
[see, e.g., the Dirichlet-Multinomial model for partitions in
\citet{Keeneretal87,McCullagh11}], but in duplicate detection we rather
expect the coreference partition to be composed by small cells.

\subsection{Missing comparisons}

The model presented in Section~\ref{ss:GeneralModel} was described
assuming that the $F$ different comparison criteria were available for
each pair of records. In practice, however, it is rather common to find
records with missing fields of information, which lead to missing
comparisons for the corresponding record pairs. If a certain field is
missing for record $i$, and this field is being used to compute
comparison data, then the vector $\bg_{ij}$, $j\neq i$, will be
incomplete, regardless of whether the field is missing for record $j$.

In order to deal with this common situation, we assume that the missing
comparisons occur at random
[MAR assumption in \citet{LittleRubin02}], and therefore we can base
our inferences on the marginal distribution of the observed comparisons
[\citet{LittleRubin02}, page 90]. The complete array of comparisons $\bG
$ can be decomposed into observed $\bG^{\mathrm{obs}}$ and missing $\bG^{\mathrm{mis}}$
comparisons; similarly, for each record pair $\bG_{ij}=(\bG
^{\mathrm{obs}}_{ij},\bG^{\mathrm{mis}}_{ij})$.
From equation \eqref{eq:PG_D}, summing over the possible missing
comparison patterns, it is easy to see that the probabilities involving
$\bG^{\mathrm{obs}}$ can be computed as
%
\begin{eqnarray}
\label{eq:PGobs_D}&& \mathbb{P} \bigl(\bG^{\mathrm{obs}}
 = \bg^{\mathrm{obs}}| \bD, \Phi
\bigr)
\nonumber
\\[-8pt]
\\[-8pt]
\nonumber
&&\qquad= \prod_{(i,j)\in\mathcal{C}} \mathbb{P}_1 \bigl(
\bg_{ij}^{\mathrm{obs}}|\Phi _1 \bigr)^{\Delta_{ij}}
\mathbb{P}_0 \bigl(\bg_{ij}^{\mathrm{obs}}|\Phi_0
\bigr)^{1-\Delta
_{ij}}\prod_{(i,j)\in\mathcal{P}-\mathcal{C}}
\mathbb{P}_0 \bigl(\bg _{ij}^{\mathrm{obs}}|
\Phi_0 \bigr),
\end{eqnarray}
where
%
\begin{equation}
\label{eq:P1obs} \mathbb{P}_1 \bigl(\bg^{\mathrm{obs}}_{ij}|
\Phi_1 \bigr)=\sum_{\bg^{\mathrm{mis}}_{ij}}\mathbb
{P}_1 \bigl(\bg^{\mathrm{obs}}_{ij},\bg^{\mathrm{mis}}_{ij}|
\Phi_1 \bigr),
\end{equation}
and we obtain an analogous expression for $\mathbb{P}_0(\bg
^{\mathrm{obs}}_{ij}|\Phi_0)$. Notice that equation \eqref{eq:PG_D} is a
particular case of equation \eqref{eq:PGobs_D} arising when all the
comparisons are complete for each record pair. In Section~\ref{s:modelindepcomps} we present a simple model under which this
approach leads to a straightforward treatment of missing comparisons.
Finally, we notice that equation \eqref{eq:PGobs_D} can be rewritten in
terms of partition\vadjust{\goodbreak} labelings as
%
\begin{eqnarray}
\label{eq:PGobs_Z}&& \mathbb{P} \bigl(\bg^{\mathrm{obs}}| \bZ, \Phi \bigr)
\nonumber\\
&&\qquad= \prod
_{(i,j)\in\mathcal{C}} \mathbb{P}_1 \bigl(
\bg_{ij}^{\mathrm{obs}}|\Phi _1 \bigr)^{I(Z_i=Z_j)}
\mathbb{P}_0 \bigl(\bg_{ij}^{\mathrm{obs}}|\Phi_0
\bigr)^{I(Z_i\neq
Z_j)}\\
&&\qquad\quad{}\times\prod_{(i,j)\in\mathcal{P}-\mathcal{C}}
\mathbb{P}_0 \bigl(\bg _{ij}^{\mathrm{obs}}|
\Phi_0 \bigr).\nonumber
\end{eqnarray}

\subsection{Conditional interpretation of the model}

Let us think about the hypothetical scenario where we know the
coreference partition for all the records except for the $i$th one. In
this case we are interested in finding the probabilities that record
$i$ refers to the different $r$ potential entities given the comparison
data, the model parameters and the partition memberships of the
remaining records, represented by an arbitrary labeling $\bZ^{(-i)}$.
Using the prior for $\bZ$ presented in Section~\ref{ss:PriorCorefMatrix}, regardless of the parametrization used for $G_1$
and $G_0$, one can show that the probability that $i$ refers to
potential entity $q$ is given by
%
\begin{eqnarray}
\label{eq:conditional}&& \mathbb{P} \bigl(Z_{i}=q | \bZ^{(-i)},
\bg^{\mathrm{obs}}, \Phi \bigr)
\nonumber
\\[-8pt]
\\[-8pt]
\nonumber
&&\qquad\propto \cases{ %
\displaystyle\prod
_{j \dvtx Z_j=q} I \bigl({(i,j)\in\mathcal{C}}
\bigr) \biggl[\frac{\mathbb{P}_1(\bg^{\mathrm{obs}}_{ij}|\Phi_1)}{\mathbb{P}_0(\bg
^{\mathrm{obs}}_{ij}|\Phi_0)} \biggr], \vspace*{2pt}\cr
\qquad \mbox{if $q$ labels  a partition cell according to $\bZ^{(-i)}$;}
\vspace*{2pt}\cr
\bigl(r-n \bigl(\bZ^{(-i)} \bigr) \bigr)^{-1},\vspace*{2pt}\cr
\qquad\mbox{otherwise.}}
\end{eqnarray}
This expression has a simple interpretation. The ratio within square
brackets in the right-hand side of equation \eqref{eq:conditional}
represents the likelihood ratio for testing the hypothesis ``\emph
{records $i$ and $j$ are coreferent}'' versus ``\emph{records $i$ and
$j$ are not coreferent},'' using the observed comparison vector $\bg
_{ij}^{\mathrm{obs}}$. If $q$ is a label in $\bZ^{(-i)}$, then the probability
that $i$ refers to entity $q$ is the product of the likelihood ratios
for all records that refer to entity $q$ according to $\bZ^{(-i)}$ (all
records $j$ such that $Z_j=q$), which is a measure of how likely record
$i$ is to be coreferent with the group of records in cell $q$. However,
if there is a record $j$ such that $Z_j=q$, but $(i,j)\notin\mathcal
{C}$, that is, $(i,j)$ was fixed as noncoreferent, then $\mathbb
{P}(Z_{i}=q | \bZ^{(-i)}, \bg^{\mathrm{obs}}, \Phi)=0$. Finally, if $q$ is a
label not in use, then record $i$ takes this label with probability
inversely proportional to the number of unused labels, which is
equivalent to saying that record $i$ gets its own label with
probability proportional to one, and the specific label is chosen
uniformly at random among the $r-n(\bZ^{(-i)})$ labels not in use.
Without being exhaustive, equation \eqref{eq:conditional} states that
if for all partition cells the products of likelihood ratios are much
smaller than one, then record $i$ will assume its own label with high
probability, but if there is a cell partition for which we obtain a
product of likelihood ratios much larger than one, then it is likely
that record $i$ gets assigned to that cell. Equation \eqref
{eq:conditional} is used in the supplementary material [\citet{SadinleAOASSupplement}] to derive a Gibbs sampler for the model
presented in the next section.

\section{A model for independent comparison fields}\label{s:modelindepcomps}

In this section we describe a simple parametrization for $G_1$ and
$G_0$, which represent the distributions of the comparison vectors
among coreferent and noncoreferent pairs, respectively. Our model
assumes that the comparison fields are independent for both coreferent
and noncoreferent records.

If comparison $\Gamma^f_{ij}$ takes $L_f+1$ values corresponding to
levels of disagreement, its distribution among coreferent records can
be modeled according to a multinomial distribution, this is
%
\begin{equation}
\label{eq:compdataMnom} \mathbb{P}_1 \bigl(\Gamma^f_{ij}=
\gamma^f_{ij}|\bmm_f \bigr)= \prod
_{l=0}^{L_f} \bigl(m_{fl}^{*}
\bigr)^{I(\gamma^f_{ij}=l)},
\end{equation}
where $\gamma^f_{ij}$ represents an observed level of disagreement,
$m_{fl}^{*}=\mathbb{P}_1(\Gamma^f_{ij}=l)$, and $\sum_{l=0}^{L_f}m_{fl}^{*}=1$. It is easy to show that these probabilities
can be rewritten as
\[
m_{fl}^{*}=\cases{ %
m_{f0}, &\quad $\mbox{if } l=0;$
\vspace*{2pt}\cr
\displaystyle m_{fl}\prod_{h<l} (1-m_{fh}), &\quad
$\mbox{if } 0<l<L_f;$
\vspace*{2pt}\cr
\displaystyle \prod_{h<L_f} (1-m_{fh}), &\quad $\mbox{if }
l=L_f;$}
\]
where $m_{f0}=\mathbb{P}_1(\Gamma^f_{ij}=0)$, and $m_{fl}=\mathbb
{P}_1(\Gamma^f_{ij}=l|\Gamma^f_{ij}>l-1)$ for $0<l<L_f$. We choose to
parameterize $G_1$ in terms of the sequential conditional probabilities
$m_{fl}$ since this facilitates prior specification, as we show in
Section~\ref{ss:priorPhi}. Using this parametrization, equation \eqref
{eq:compdataMnom} can be reexpressed as
\[
\mathbb{P}_1 \bigl(\bG^f_{ij}=
\bg^f_{ij}|\bmm_f \bigr) = \prod
_{l=0}^{L_f-1} m_{fl}^{I(\gamma^{f}_{ij}=l)}(1-m_{fl})^{I(\gamma
^{f}_{ij}>l)},
\]
where $\bmm_f=(m_{f0},\ldots,m_{f,L_f-1})$. Following an analogous
construction of the distribution of $\Gamma^f_{ij}$ among noncoreferent
pairs, we obtain
\[
\mathbb{P}_0 \bigl(\bG^f_{ij}=
\bg^f_{ij}|\bu_f \bigr) = \prod
_{l=0}^{L_f-1} u_{fl}^{I(\gamma^{f}_{ij}=l)}(1-u_{fl})^{I(\gamma^{f}_{ij}>l)},
\]
where\vspace*{1pt} $u_{f0}=\mathbb{P}_0(\Gamma^f_{ij}=0)$, $u_{fl}=\mathbb
{P}_0(\Gamma^f_{ij}=l|\Gamma^f_{ij}>l-1)$ for $0<l<L_f$, and $\bu
_f=(u_{f0},\ldots,u_{f,L_f-1})$. Notice that if $L_f=1$, that is, if
comparison $f$ is binary, we obtain the traditional model used in
record linkage for binary comparisons [e.g., \citet{Winkler88,Jaro89}].

\subsection{Missing comparisons and conditional independence}

The assumptions of the comparison fields being conditionally
independent (CI), along with being missing at random (MAR), make it
straightforward to deal with missing comparisons. In fact, under these
assumptions, equation \eqref{eq:P1obs} can be written as
%
\begin{equation}
\label{eq:P1missing} \mathbb{P}_1 \bigl(\bg^{\mathrm{obs}}_{ij}|
\Phi_1 \bigr) = \prod_{f=1}^{F}
\Biggl[\prod_{l=0}^{L_f-1} m_{fl}^{I(\gamma
^{f}_{ij}=l)}(1-m_{fl})^{I(\gamma^{f}_{ij}>l)}
\Biggr]^{I_{\mathrm{obs}}(\gamma_{ij}^f)},
\end{equation}
where $I_{\mathrm{obs}}(\cdot)$ is one if its argument is observed, and zero if
it is missing, and $\Phi_1=(\bmm_1,\ldots,\bmm_F)$. Similarly,
%
\begin{equation}
\label{eq:P0missing} \mathbb{P}_0 \bigl(\bg^{\mathrm{obs}}_{ij}|
\Phi_0 \bigr) = \prod_{f=1}^{F}
\Biggl[\prod_{l=0}^{L_f-1} u_{fl}^{I(\gamma
^{f}_{ij}=l)}(1-u_{fl})^{I(\gamma^{f}_{ij}>l)}
\Biggr]^{I_{\mathrm{obs}}(\gamma_{ij}^f)},
\end{equation}
where $\Phi_0=(\bu_1,\ldots,\bu_F)$. Equations \eqref{eq:P1missing} and
\eqref{eq:P0missing} indicate that the combination of the CI and MAR
assumptions allow us to ignore the comparisons that are not observed
and yet model the observed comparisons in a simple fashion.

Under the CI assumption we can write the likelihood for $\bZ$ and $\Phi
$ as
\[
\mathcal{L}(\bZ,\Phi|\bg_{\mathrm{obs}}) = \prod_{f=1}^F
\mathcal{L} \bigl(\bZ,\Phi _f|\bg^{f}_{\mathrm{obs}} \bigr),
\]
where $\Phi_f = (\bmm_f, \bu_f)$, and
\[
\mathcal{L} \bigl(\bZ,\Phi_f|\bg^{f}_{\mathrm{obs}} \bigr)
= \prod_{l=0}^{L_f-1} m_{fl}^{a^1_{fl}(\bZ)}(1-m_{fl})^{\sum_{h>l}
a^1_{fh}(\bZ)}u_{fl}^{a^0_{fl}(\bZ)}(1-u_{fl})^{\sum_{h>l}a^0_{fh}(\bZ)},
\]
where
\begin{eqnarray*}
a^1_{fl}(\bZ) &=& \sum_{(i,j)\in\mathcal{C}}I_{\mathrm{obs}}
\bigl(\bg_{ij}^f \bigr)I \bigl(\gamma _{ij}^{f}=l
\bigr)I(Z_i=Z_j),
\\
a^0_{fl}(\bZ) &=&\sum_{(i,j)\in\mathcal{C}}I_{\mathrm{obs}}
\bigl(\bg_{ij}^f \bigr)I \bigl(\gamma _{ij}^{f}=l
\bigr)I(Z_i\neq Z_j)+\sum_{(i,j)\in\mathcal{P}-\mathcal
{C}}I_{\mathrm{obs}}
\bigl(\bg_{ij}^f \bigr)I \bigl(\gamma_{ij}^{f}=l
\bigr).
\end{eqnarray*}
For a given matrix of memberships $\bZ$, $a^1_{fl}(\bZ)$ and
$a^0_{fl}(\bZ)$ represent the number of coreferent and noncoreferent
records disagreeing at level $l$ for observed comparison $f$.

Although our main interest is to make inferences on the coreference
matrix $\bD$, a fully Bayesian approach requires the specification of
priors for the parameters $\Phi$ as well.

\subsection{Prior specification for the model parameters}\label{ss:priorPhi}

We now explain our selection of the priors for $m_{fl}$ and $u_{fl}$,
$l=0,\ldots,L_f-1$. The first parameter that we focus on is
$m_{f0}=\mathbb{P}_1(\Gamma^f_{ij}=0)$, which represents the
probability of observing the level zero of disagreement in the
comparison $f$ among coreferent records. This level represents no
disagreement or a high degree of agreement, so if we believe that field
$f$ contains no error, $m_{f0}$ should be, a priori, a point mass at
one, but as the error in field $f$ increases, the mass of $m_{f0}$'s
prior should move away from one. We therefore take a priori $m_{f0}$ to
be in some interval $[\lambda_{f0},1]$ with probability one, for some
$0<\lambda_{f0}<1$. If we believe that the field used to compute
comparison $f$ is fairly accurate, then we should set the threshold
$\lambda_{f0}$ to be close to one. On the other hand, the more errors
we believe a field contains, the lower the value for $\lambda_{f0}$
that we should set. The prior distribution for $m_{f0}$ can be taken in
general as Beta$(\alpha_{f0}^1,\beta_{f0}^1)$, truncated to the
interval $[\lambda_{f0},1]$, which we denote as $\operatorname{TBeta}(\alpha
_{f0}^1,\beta_{f0}^1,\lambda_{f0},1)$.

The parameter $m_{f1}=\mathbb{P}_1(\Gamma^f_{ij}=1|\Gamma^f_{ij}>0)$
represents the probability of observing level one of disagreement in
the comparison $f$, among coreferent record pairs with disagreement
larger than the one captured by the level zero. Depending on the
construction of the disagreement levels, and if the number of levels is
greater than two, we can think of level one of disagreement as some
mild disagreement and, therefore, if we expect the amount of error to
be relatively small, $m_{f1}$ should be concentrated around values
close to one. Following a similar reasoning as for $m_{f0}$, we take
the prior of $m_{f1}$ as $\operatorname{TBeta}(\alpha_{f1}^1,\beta
_{f1}^1,\lambda_{f1},1)$, where we can set the hyperparameters of this
distribution, especially $\lambda_{f1}$, according to the amount of
error that we expect field $f$ to contain.

We can continue the previous reasoning to specify the prior
distribution of the remaining parameters $m_{fl}=\mathbb{P}_1(\Gamma
^f_{ij}=l|\Gamma^f_{ij}>l-1)$, $l=2,\ldots,L_f-1$. In general, we can
take the prior of $m_{fl}$ as $\operatorname{TBeta}(\alpha_{fl}^1,\beta
_{fl}^1,\lambda_{fl},1)$, where the truncation points $\lambda_{fl}$
change according to the way the disagreement levels were constructed
and the amount of error expected a priori in each field. Notice,
however, that if we believe that a field may be too erroneous, it may
be better to exclude it from the duplicate detection process since its
inclusion can potentially harm the results [\citet{SadinleFienberg13}
explore this issue in the multiple record linkage context]. For
simplicity, in this article we set $\alpha_{fl}^1=\beta_{fl}^1=1$, for
all fields $f$ and levels~$l$, that is, we take $m_{fl}\sim\operatorname
{Uniform}(\lambda_{fl},1)$, and so we only need to choose the $\lambda_{fl}$'s.

The probabilities $u_{fl}=\mathbb{P}_0(\Gamma^f_{ij}=l|\Gamma
^f_{ij}>l-1)$ among noncoreferent records may have quite different
distributions depending on the fields used to compute the comparisons.
For instance, if a nominal field contains a highly frequent category,
then the probability of agreement will be high even for noncoreferent
records. On the other hand, if a field is almost a unique identifier of
the entities, then the probability of agreement will be small among
noncoreferent records. We therefore simply take $u_{fl}\sim\operatorname
{Uniform}(0,1)$ for all fields and levels of disagreement, although in
general we could take $u_{fl}\sim\operatorname{Beta}(\alpha_{fl}^0,\beta
_{fl}^0)$, for some hyperparameters $\alpha_{fl}^0$ and $\beta_{fl}^0$
if prior information was available.

\subsection{Bayesian inference via Gibbs sampler}\label{ss:Gibbs}

In the supplementary material [\citet{SadinleAOASSupplement}] we present
a Gibbs sampler to explore the joint posterior of $\bZ$ and $\Phi$
given the observed comparison data $\bg_{\mathrm{obs}}$, for the likelihood
obtained from equations \eqref{eq:PGobs_Z}, \eqref{eq:P1missing} and
\eqref{eq:P0missing}, and the priors presented in the previous
subsection. The supplementary material also contains a brief discussion
on point estimation of the coreference partition.

\begin{table}
\caption{Illustrative example: Different sets of records may be
considered as coreferent in different contexts}\label{t:toyexample}
\begin{tabular*}{\textwidth}{@{\extracolsep{\fill}}lcccccc@{}}
\hline
\textbf{Record} & \textbf{Given name} & \textbf{Family name} &
\textbf{Year} & \textbf{Month} & \textbf{Day} & \textbf{Municipality} \\
\hline
1. & JOSE & FLORES & 1981 & 1 & 29 & A \\
2. & JOSE & FLORES & 1981 & 2 & NA & A \\
3. & JOSE & FLORES & 1981 & 3 & 20 & A \\
4. & JULIAN ANDRES & RAMOS ROJAS & 1986 & 8 & \phantom{0}5 & B \\
5. & JILIAM & RMAOS & 1986 & 8 & \phantom{0}5 & B \\
\hline
\end{tabular*}
\end{table}
%

\subsection{An illustrative example}\label{ss:toyexample}

Table~\ref{t:toyexample} presents a small example to illustrate
different situations where different sets of records may be considered
as coreferent depending on how reliable we believe the fields are. We
explore the results of our duplicate detection method under different
scenarios where these data could have arisen, which is why we do not
yet specify what the fields year, month, day, and municipality refer
to. This example was inspired by the data file that we study in Section~\ref{s:UNTRCapplication}, where we have to compare Hispanic names.
Full Hispanic names are usually composed by four pieces, two
corresponding to given name and two to family name. In practice,
however, Hispanic people do not always use their full given and family
names. For example, someone whose full given name is \textit{JULIAN
ANDRES} may be simply known as \textit{JULIAN} or as \textit{ANDRES} in his
social circle. This phenomenon makes it particularly challenging to
compare Hispanic names, for example, it has been reported to cause
problems when tracking citations of Hispanic authors [\citet{RuizPerezetal02,FernandezGarcia03}].

Records 1, 2 and 3 in Table~\ref{t:toyexample} represent an example
where pairwise decisions on the coreference statuses of record pairs
may not be transitive. In this example, records 1, 2 and 3 agree in all
the fields except for month and day. Records 1 and 2 disagree by one
month, as well as records 2 and 3, but the comparison for the field day
for those two pairs is missing. Notice also that records 1 and 3
disagree by two months and have a strong disagreement in the field day.
In this situation, a~method taking pairwise decisions, or even a human
taking decisions for one pair of records at a time, may decide that
records 1 and 2 are coreferent, as well as records 2 and 3, since those
pairs are fairly similar, but may decide that records 1 and 3 are not
coreferent, since this pair has more disagreements.
Table~\ref{t:toyexample} also presents records 4 and 5, which agree in all of
their information, except for given and family name. Record 5 could
refer to the same person as record 4, since this name is simply missing
the second pieces of given and family name, which is common for
Hispanic names, and the remaining disagreements could be typographical
errors. The decision of whether to declare records 4 and 5 as
coreferent will depend on the levels of error that we believe the
fields given and family name may contain. Below we show how the
proposed method deals with the uncertainty of these situations under
different scenarios.

Let us think of two different scenarios from where the records in Table~\ref{t:toyexample} could have arisen. In the first scenario, inspired
by the application presented in Section~\ref{s:UNTRCapplication}, each
record refers to a person who was killed during a war, and the data
were reported by witnesses many years after the events occurred. In
this scenario, year, month, day and municipality correspond to the date
and location of the killing as reported by the witnesses. Under this
scenario we expect to have reporting errors in the names of the victim
and in the date and place of the killings, since different witnesses
may have different memories of the victims and the events. In the
second scenario, the records in Table~\ref{t:toyexample} come from tax
forms, and the information was self-reported. In this case, year,
month, day and municipality correspond to date and place of birth. In
this case we may expect the levels of error in all fields to be much
smaller compared to the first scenario, since it is quite unlikely for
one person to misreport her information, at least unintentionally.

\begin{table}
\caption{Construction of levels of disagreement for the example in
Table \protect\ref{t:toyexample}}\label{t:toyexample_compdata}
\begin{tabular*}{\textwidth}{@{\extracolsep{\fill}}lccccc@{}}
\hline
& & \multicolumn{4}{c@{}}{\textbf{Levels of disagreement}}\\[-4pt]
& & \multicolumn{4}{c@{}}{\hrulefill}\\
\textbf{Field} & \textbf{Similarity measure} & $\bolds{0}$ & $\bolds{1}$ &
$\bolds{2}$ & \multicolumn{1}{c@{}}{$\bolds{3}$} \\
\hline
Given name & Modified Levenshtein & 0 & $(0,0.25]$ & $(0.25,0.5]$ &
$(0.5,1]$ \\
Family name & Modified Levenshtein & 0 & $(0,0.25]$ & $(0.25,0.5]$ &
$(0.5,1]$ \\
Year & Absolute difference & 0 & 1 & 2--3 & $4+$ \\
Month & Absolute difference & 0 & 1 & 2--3 & $4+$ \\
Day & Absolute difference & 0 & 1--2 & 3--7 & $8+$ \\
Municipality & Binary comparison & Agree & Disagree &&\\
\hline
\end{tabular*}
\end{table}

In Table~\ref{t:toyexample_compdata} we show a summary of how we
construct disagreement levels in this example. We compare all the
record pairs since there are only 10 of them and use a modification of
the \emph{Levenshtein edit distance} to compare names. The Levenshtein
edit distance between two strings is the minimum number of deletions,
insertions or replacements that we need to transform one string into
the other. The modification that we use simply accounts for the fact
that Hispanic names may have missing pieces. Basically, if name $V$
contains one token and name $W$ contains two tokens, we take the
minimum of the Levenshtein distances between the token of name $V$ and
each token of name $W$ and, finally, we transform this measure to the
0--1 interval. In this scale, 0 means total agreement (up to missing
tokens) and 1 means extreme disagreement. We refer the reader to the
supplementary material [\citet{SadinleAOASSupplement}] for details on
our comparisons of Hispanic names. The intervals that we choose to
construct the disagreement levels (except for municipality) correspond
to what we consider as no disagreement, mild disagreement, moderate
disagreement and extreme disagreement. In this example the field
municipality is taken as a nominal variable, and so we compare it in a
binary fashion.

\begin{figure}
\tabcolsep=0pt
{\fontsize{9}{11}\selectfont{
\begin{tabular*}{\textwidth}{@{\extracolsep{4in minus 4in}}lccc@{}}
\hline
& \multicolumn{2}{c}{\textbf{Prior truncation points for} $\bolds{\{m_{fl}\}}$}&\\[-6pt]
&\multicolumn{2}{c}{\rule{125pt}{0.5pt}}&\\
& \multicolumn{1}{c}{\textbf{Given and}} &\multicolumn{1}{c}{\textbf{Day and}}&
\\
& \multicolumn{1}{c}{\textbf{family names}}&\multicolumn{1}{c}{\textbf{Month}}&
\multicolumn{1}{c@{}}{\textbf{Posterior frequencies}}\\
\hline
1. & 0.85 & 0.85 &
\includegraphics{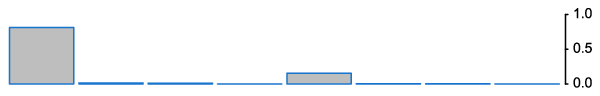}
 \\
2. & 0.85 & 0.95 &
\includegraphics{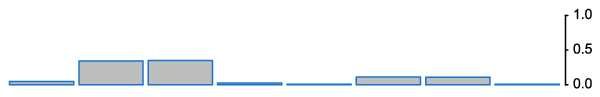}
 \\
3. & 0.95 & 0.85 &
\includegraphics{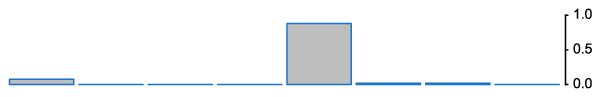}
 \\
4. & 0.95 & 0.95 &
\includegraphics{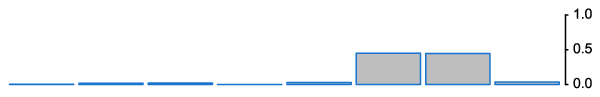}
\\[3pt]
\multicolumn{3}{@{}c}{Coreference matrices} &\multicolumn{1}{l}{
\includegraphics{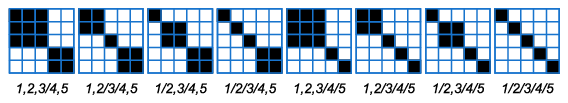}
}\\[2pt]
\hline
\end{tabular*}}}
\caption{Posterior distributions of the coreference partition for the
records in Table~\protect\ref{t:toyexample}, for different sets of priors
corresponding to different contexts. Prior truncation points for Year
and Municipality parameters are set at 0.95 for all cases. Posterior
frequencies are obtained from 9000 iterations of a Gibbs sampler. The
eight partitions presented here concentrate 100\% of the posterior
frequencies in each case. The coreference matrices depicted here have
black entries representing ones and white entries representing zeroes.
}\label{f:pp}
\end{figure}

To implement the proposed method for duplicate detection, we need to
choose the prior truncation points of the parameters $m_{fl}$. For the
sake of simplicity, we suppose that our prior beliefs about each field
of information can be classified in two categories: either the field is
nearly accurate or it is inaccurate. If field $f$ is nearly accurate,
we take the prior truncation points for all the parameters related to
this field (all $m_{fl}$, $l=0,\ldots,L_f-1$) as 0.95, whereas if field
$f$ is inaccurate, these prior truncation points are set to 0.85. For
simplicity, we fix the prior truncation points for year and
municipality parameters at 0.95 for all of the data collection
scenarios presented here. For the remaining parameters, in the war
scenario we expect the fields to contain considerable amounts of error,
and so the prior truncation points for those parameters are set equal
to 0.85 (case 1 of Figure~\ref{f:pp}); for the taxes scenario the prior
truncation points are set equal to 0.95 since a priori we expect errors
to be rare (case 4 of Figure~\ref{f:pp}). We also explore two
intermediate cases that fall between the previous two extreme
scenarios, where we consider day and month to be nearly accurate, but
given and family names to be inaccurate (case 2 of Figure~\ref{f:pp})
and vice versa (case 3 of Figure~\ref{f:pp}).

For each set of priors we run 10,000 iterations of the Gibbs sampler
presented in the supplementary material [\citet{SadinleAOASSupplement}],
and in each case we discard 1000 iterations as burn-in. Figure~\ref{f:pp} presents the posterior frequencies of the eight partitions that
appear in the posterior samples. Although a file with five records can
be partitioned in 52 ways (the 5th Bell number), the eight partitions
presented in Figure~\ref{f:pp} concentrate 100\% of the posterior
frequencies in each case.

From Figure~\ref{f:pp} we can see that for case 1, that is, when given
and family names, and day and month are inaccurate, the posterior
distribution is mostly concentrated in partition $1,2,3/4,5$, that is,
records 1, 2 and 3 are assigned to one entity and records 4 and 5 to
another; this result is coherent with our priors, which indicated that
the fields were potentially inaccurate, and therefore the disagreements
between fields are not taken as strong evidence of the records being
noncoreferent. In case 2, given and family names are thought to be
inaccurate, whereas day and month are considered to be fairly accurate;
in this case the strong disagreements between records 1 and 3 become
important evidence of them not being coreferent, but since record pairs
1 and 2, and 2 and 3 are quite similar, the partitions $1,2/3/4,5$ and
$1/2,3/4,5$ get equal posterior probability. In case 3, we present a
scenario where day and month are thought to be inaccurate, but given
and family names are believed to be accurate, and therefore the
posterior gets almost completely concentrated in the partition $1,2,3/4/5,$
that is, compared to case 1, disagreements in given and family names
become more important for distinguishing noncoreferent records, and
therefore records 4 and 5 are probably noncoreferent. Finally, in case
4, all the fields are considered as accurate, and therefore the
partitions where records 4 and 5 are coreferent become unlikely a
posteriori, as well as the partitions where records 1 and 3 are
clustered together. Since records 1 and 2 are quite similar, as well as
records 2 and 3, but records 1 and 3 have strong disagreements, the
posterior assigns equal probability to the partitions $1,2/3/4/5$ and
$1/2,3/4/5,$ which accounts properly for the uncertainty of deciding
whether records 1 and 2, or records 2 and 3 are coreferent.

Finally, it is important to emphasize that although in this example it
seems that the priors of the $m_{fl}$ parameters completely determine
the posterior of $\bD$, both the $m_{fl}$ and $u_{fl}$ parameters
influence the evolution of the memberships $\bZ$ in the Gibbs sampler
(see the supplementary material [\citet{SadinleAOASSupplement}]). In
particular, if these five records were contained in a larger file,\vadjust{\goodbreak} the
resolution of their coreference statuses would depend on the
distribution of the comparison data for the complete file, since, for
instance, the distributions of the $u_{fl}$ parameters are heavily
influenced by the observed frequencies of the corresponding levels of
disagreement.

\subsection{A simulation study}\label{ss:simAustralia}

We now present a simulation study to explore the performance of the
proposed methodology under different scenarios of measurement error.
Peter Christen and his collaborators [\citet{Christen05,ChristenPudjijono09,ChristenVatsalan13}] developed a
sophisticated data generation and corruption tool to create synthetic
data sets containing various types of fields. This tool, written in
Python, can include dependencies between fields, permits the generation
of different types of errors, and can be easily adapted to generate
additional fields that are not included in the default settings.

We now describe the characteristics of the data files used in the
simulation. We consider files having either five or seven fields of
information. The synthetic files involving five fields include the
following: gender, given name, family name, age, and occupation. The
files with seven fields additionally include postal code and phone
number. The fields gender and given name are sampled jointly from a
table that contains frequencies of given names per gender, and
therefore popular given names appear with higher probability in the
synthetic data sets. Family name and postal codes are generated
independently from additional frequency tables. The three tables
mentioned so far were compiled by Christen and his collaborators using
public sources from Australia. Phone numbers are randomly generated
following the Australian format which consists of a two-digit area code
and an eight-digit number made of two blocks of four digits. The
previously described fields were included in the default configuration
of Christen's generator. In addition, age and occupation are jointly
sampled from a contingency table that serves as an estimate of the
distribution of age and occupation in Australia. This table was
obtained from the webpage of the Australian Bureau of Statistics, and
it contains eight categories of occupation and eight age intervals.

The generator first creates a number of original records which are
later used to create distorted duplicates. The duplicates are allocated
by randomly selecting an original record and assigning a random number
of duplicates to it. The number of duplicates is generated according to
a Poisson(1) truncated to the interval $[1, 5]$. Each duplicate has a
fixed number of erroneous fields which are allocated uniformly at
random, and each field contains maximum two errors. The types of errors
are selected uniformly at random from a set of possibilities which vary
from field to field, as summarized in Table~\ref{t:errors_simulationAustralia}. In this table, missing values means
that the value of the field becomes missing; edit errors represent
random insertions, deletions or substitutions of characters in the
string; OCR errors happen typically when a document has been digitized
using optical character recognition; keyboard errors use a keyboard
layout to simulate typing errors; phonetic errors are simulated using a
list of predefined phonetic rules; and finally, misspelling errors are
generated by randomly selecting one of possibly many known misspellings
of a family name. For further details on the generation of these types
of errors, see \citet{ChristenPudjijono09} and \citet{ChristenVatsalan13}.

\begin{table}
\caption{Types of errors per field in the simulation study of Section
\protect\ref{ss:simAustralia}}\label{t:errors_simulationAustralia}
\begin{tabular*}{\textwidth}{@{\extracolsep{\fill}}lcccccc@{}}
\hline
& \multicolumn{6}{c@{}}{\textbf{Type of error}}\\[-4pt]
& \multicolumn{6}{c@{}}{\hrulefill}\\
\textbf{Field} & \textbf{Missing values} & \textbf{Edits} & \textbf{OCR} &
\textbf{Keyboard} & \textbf{Phonetic} &
\textbf{Misspelling} \\
\hline
Given name & & $\checkmark$ & $\checkmark$ & $\checkmark$ & $\checkmark
$ & \\
Family name & & $\checkmark$ & $\checkmark$ & $\checkmark$ &
$\checkmark$ & $\checkmark$ \\
Age interval & $\checkmark$ & &&&& \\
Gender & $\checkmark$ & &&&& \\
Occupation & $\checkmark$ & &&&& \\[3pt]
Phone number & $\checkmark$ & $\checkmark$ & $\checkmark$ & $\checkmark
$ && \\
Postal code & $\checkmark$ & $\checkmark$ & $\checkmark$ & $\checkmark
$ && \\
\hline
\end{tabular*}
\end{table}

\begin{table}[b]
\caption{Construction of levels of disagreement for the simulation
study of Section \protect\ref{ss:simAustralia}}\label{t:simAustralia_compdata}
\begin{tabular*}{\textwidth}{@{\extracolsep{\fill}}lccccc@{}}
\hline
& & \multicolumn{4}{c@{}}{\textbf{Levels of disagreement}}\\[-4pt]
& & \multicolumn{4}{c@{}}{\hrulefill}\\
\textbf{Field} & \textbf{Similarity measure} & $\bolds{0}$ & $\bolds{1}$ &
$\bolds{2}$ & \multicolumn{1}{c@{}}{$\bolds{3}$} \\
\hline
Given name & Levenshtein & 0 & $(0,0.25]$ & $(0.25,0.5]$ & $(0.5,1]$ \\
Family name & Levenshtein & 0 & $(0,0.25]$ & $(0.25,0.5]$ & $(0.5,1]$
\\
Age interval & Binary comparison & Agree & Disagree &&\\
Gender & Binary comparison & Agree & Disagree &&\\
Occupation & Binary comparison & Agree & Disagree &&\\[3pt]
Phone number & Levenshtein & 0 & $(0,0.25]$ & $(0.25,0.5]$ & $(0.5,1]$
\\
Postal code & Levenshtein & 0 & $(0,0.25]$ & $(0.25,0.5]$ & $(0.5,1]$
\\
\hline
\end{tabular*}
\end{table}

In the simulation presented here, each synthetic data set is composed
of 450 original records and 50 duplicates. To explore the performance
of the method as a function of the amount of error in the data file, we
generate 100 five-field and 100 seven-field synthetic data sets for
each of three levels of error, which correspond to the number of
erroneous fields per duplicate being one, three and five. For each
file, comparison data were created as indicated in Table~\ref{t:simAustralia_compdata}. For these files we model all pairs, so
$|\mathcal{P}|={500\choose 2}$, and the record pairs having the level
three of disagreement in either given or family name were fixed as
noncoreferent, so these pairs constitute the set $\mathcal{P}-\mathcal
{C}$, as explained in Section~\ref{ss:complexity}. Our model is then
applied under three different sets of priors. For simplicity, each set
of priors has the same prior truncation point for all the $m_{fl}$
parameters, although in practice the priors should be chosen carefully
based on knowledge of the potential amounts of error in the file. The
prior truncation points are 0.5, 0.8 and 0.95, which correspond to one
scenario where we believe the amount of error in the file to be
extremely large, one where we believe it to be moderate, and one where
we are optimistic and believe the amount of error to be very limited.
For each data set, and for each set of priors, we ran 10,000 iterations
of the Gibbs sampler and discarded the first 1000 as burn-in. The
average runtime using an implementation in R [\citet{R13}] with parts
written in C
language was of 24.5 seconds per file, including the computation of the
comparison data, on a laptop with a 2.80 GHz processor. Before starting
the complete simulation study, we obtained some longer chains for some
data sets and for all priors, and we could check that 9000 iterations
provided roughly the same frequencies of partitions as when we ran
longer chains.

For each data file, and each set of priors, we obtain a sample of
partitions which approximate the posterior distribution of the
coreference partition. We can assess how good each partition is in
terms of classifying pairs of records as coreferent and noncoreferent.
Two records $i$ and $j$ are coreferent according to a partition $\bD'$
if both belong to the same cell of the partition, that is, $\Delta
_{ij}'=1$. Given $\bD'$ and the true partition $\bD^*$, let $b_{11}(\bD
',\bD^*)=\sum_{i<j}\Delta_{ij}'\Delta_{ij}^*$ be the number of record
pairs that are coreferent in both partitions, and $b_{10}(\bD',\bD
^*)=\sum_{i<j}\Delta_{ij}'(1-\Delta_{ij}^*)$ and $b_{01}(\bD',\bD
^*)=\sum_{i<j}(1-\Delta_{ij}')\Delta_{ij}^*$ be the number of record
pairs that are coreferent in one partition but not in the other. Given
that $\bD^*$ is the true partition, the \emph{recall} of $\bD'$ is
defined as $b_{11}(\bD',\bD^*)/ (b_{11}(\bD',\bD^*)+b_{01}(\bD',\bD
^*) )$, whereas the \emph{precision} of $\bD'$ is $b_{11}(\bD',\bD
^*)/ (b_{11}(\bD',\bD^*)+b_{10}(\bD',\bD^*) )$. The recall of
a partition $\bD'$ measures the proportion of truly coreferent pairs
that are classified correctly by $\bD'$, whereas the precision of $\bD
'$ measures the proportion of pairs declared as coreferent by $\bD'$
that are truly coreferent. These two measures are preferred for
evaluating performance in duplicate detection and record linkage
problems, where the set of noncoreferent pairs is much bigger than the
set of coreferent pairs, and therefore traditional measures of
performance in classification, such as the misclassification rate and
the true negative rate, are misleading [\citet{ChristenBook}, page 165].

\begin{figure}

\includegraphics{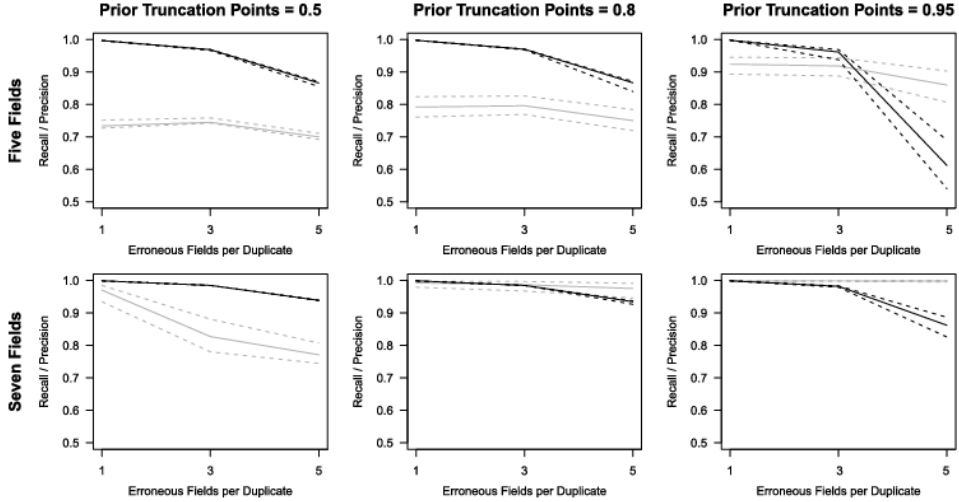}

\caption{Performance of the proposed methodology in the simulation of
Section \protect\ref{ss:simAustralia}. Black lines refer to recall, gray lines
to precision, solid lines show average medians, and dashed lines show
average first and 99th percentiles.}
\label{f:AustraliaSim}
\end{figure}

The results of the simulation are presented in Figure~\ref{f:AustraliaSim}, where the rows of panels correspond to different
number of fields and the columns to different priors. Notice that for
each data set and each set of priors we obtain a distribution of recall
and precision measures, since both of these measures are computed for
each partition in the posterior sample. Therefore, we compute the
median, the first and 99th percentile of each measure for each data set
and each set of priors, and average over all the 100 results
corresponding to each level of error, each number of fields and each
prior. In each panel of Figure~\ref{f:AustraliaSim} black lines refer
to recall, gray lines to precision, solid lines show average medians,
and dashed lines show average first and 99th percentiles.

We can see that the performance of the method depends greatly on the
amount of identifying information contained in the files (number and
type of fields) and the interplay between our prior beliefs and the
real amount of error. As we would naturally expect, our ability to
obtain results with high precision will depend on the amount\vadjust{\goodbreak} of
identifying information contained in the files, that is, in general we
will tend to obtain large proportions of false coreferent pairs
whenever we have a small number of fields (see first row of Figure~\ref{f:AustraliaSim}). For the five-field data files the precision of the
method is generally sensitive to prior specification, whereas the
recall is somewhat insensitive except for when the amount of error is
large but we believe it to be small (see upper right panel), in which
case we obtain a very poor recall, which means that a large proportion
of truly coreferent pairs will not be detected. For files with seven
fields, if the amount of error is small, then both recall and precision
are somewhat insensitive to the choice of the prior truncation points,
as long as the prior is not overly pessimistic in terms of the expected
amount of error, in which case the precision deteriorates (see bottom
left panel). This indicates that when there are not many errors, it is
easy to identify most truly coreferent pairs, but if our priors are
overly pessimistic, indicating that the amount of error is potentially
much larger than what it really is, then we will end up obtaining many
false coreferent pairs.

Although for some scenarios it is possible to obtain results that are
both good and not too sensitive to prior specification, the general
performance of the method can be seen in terms of a trade-off between
recall and precision: if the priors indicate that the amount of error
may be too small when it is actually large, then we may wind up missing
too many true coreferent pairs; if the priors indicate that the amount
of error may be too large when it is actually small, then we may end up
having too many false coreferent pairs.

The results of this simulation study provide us with some guidance for
the application presented in the next section, where the data file we
work with contains a small number of fields, and we believe its levels
of error to be intermediate.

\section{Detecting killings multiply reported to the U.N. Truth
Commission for El Salvador}\label{s:UNTRCapplication}

Unfortunately, the list of homicides obtained by the UNTC was never
made available in electronic form and was publicly available only as
photocopies as of 2007 [\citet{Amelia11}]. As part of her Ph.D. thesis,
Amelia Hoover Green utilized Optical Character Recognition (OCR)
technology, along with data cleaning and standardization, to transfer
those scanned lists into spreadsheet format. The digitized lists
therefore contained OCR errors that were corrected by hand as part of
the current project.

We now describe how we use the proposed methodology to find duplicated
homicide records in the UNTC database. The fields that we use are given
name, family name, date of death (year, month and day) and municipality
of death, similarly as in the example of Section~\ref{ss:toyexample}.
In this article, a valid homicide report is defined as a record in the
data file that specifies given and family name of the victim, which
leads to a data file containing 5395 records. We believe that no
single field in this file is free of error, and therefore we do not use
traditional blocking, as it may lead to miss many truly coreferent
pairs. There are, however, some disagreements between pairs of records
that make us confident about their noncoreference statuses.

\subsection{Filtering trivial noncoreferent record pairs, and
comparison data}

We consider it reasonable to assume that two reports correspond to
different homicides whenever their recorded municipalities have names
with no overlap and are not geographical neighbors. This approach takes
into account the fact that some homicides occurring near the boundary
of two municipalities may get reported in the wrong, although
neighboring, municipality. Another source of error occurs when a
municipality gets wrongly coded due to multiple municipalities having
similar names. Although the testimonies were collected in different
regions of El Salvador, they were digitized in a central location and,
therefore, if, for example, a~report indicated simply \textit{San
Francisco} as the municipality where a killing occurred, the clerks who
entered the data could have potentially assigned the wrong municipality
code to this report, given that there are six different municipalities
in El Salvador that include those two tokens, for example, \textit{San
Francisco Moraz\'an}, \textit{San Francisco Lempa}, among others. We
therefore only fully compare record pairs that either have the same
municipality, neighboring municipalities, municipalities with names
that overlap by at least one token (ignoring the common tokens \textit
{San, Santa, Santo, La, El, Las, Los, Del, De}), or for which the
municipality is missing. The set of pairs that meet any of the previous
criteria constitute the set $\mathcal{P}$ introduced in Section~\ref{ss:complexity}, and the remaining pairs are fixed as noncoreferent. By
using this approach we only need to fully compare around 12\% of the
${5395\choose 2}=14\mbox{,}550\mbox{,}315$ possible record pairs.

We construct the comparison data in the same way as in the illustrative
example of Section~\ref{ss:toyexample}, as summarized in Table~\ref{t:toyexample_compdata}. Given and family names were standardized and
compared as described in the supplementary material [\citet{SadinleAOASSupplement}]. The record pairs having the level three of
disagreement in either given or family name, or in year and month, were
fixed as noncoreferent (these are the pairs in the set $\mathcal
{P}-\mathcal{C}$ introduced in Section~\ref{ss:complexity}). After this
step, the number of pairs on which we still need to take decisions
reduces to only $|\mathcal{C}|=759$, which involve only 1035 records.

\subsection{Prior specification}

Following the general guidelines presented in Section~\ref{ss:priorPhi}, we use uniform priors on $[0,1]$ for all the $u_{fl}$
parameters, $f\in\{\mbox{Given name, Family name, Year, Month, Day,
Municipality}\}$, $l=0,\ldots,L_f-1$. For the $m_{fl}$ parameters, we
use flat priors in the intervals $[\lambda_{fl},1]$ for the truncation
points $\lambda_{fl}$ given in Table~\ref{t:prior_ElSalvador}.
These priors indicate our belief that coreferent pairs are very likely
to have exact agreements, although we still expect a considerable
amount of error in the fields. For example, the probability of exact
agreement in the field year of death for coreferent pairs
[$m_{\mathrm{Year},0}=\mathbb{P}_1(\Gamma^{\mathrm{Year}}_{ij}=0)$] is set to be at least
0.85 (i.e., $\lambda_{\mathrm{Year},0}=0.85$), which indicates that we expect a
pair of coreferent records to agree exactly on year of death with high
probability, but we still think that the amount of error could go up to
15\%. The remaining $\lambda_{f0}$ truncation points have similar
interpretations.

\begin{table}
\caption{Prior truncation points $\lambda_{fl}$ for the $m_{fl}$
parameters in the detection of duplicate homicide records in the UNTC
data file}\label{t:prior_ElSalvador}
\begin{tabular*}{\textwidth}{@{\extracolsep{\fill}}lcccccc@{}}
\hline
& \multicolumn{6}{c@{}}{\textbf{Field ($\bolds{f}$)}}\\[-4pt]
& \multicolumn{6}{c@{}}{\hrulefill}\\
$\bolds{l}$ & \textbf{Given name} & \textbf{Family name} &
\textbf{Year} & \textbf{Month} & \textbf{Day} & \multicolumn{1}{c@{}}{\textbf{Municipality}} \\
\hline
0 &0.85 & 0.85 & 0.85 & 0.85 & 0.70 & 0.85\\
1 &0.90 & 0.90 & 0.90 & 0.90 & 0.70 & -- \\
2 &0.99 & 0.99 & 0.99 & 0.99 & 0.70 & -- \\
\hline
\end{tabular*}
\end{table}

The truncation points for the remaining parameters reflect our belief
on the fields' error structure. We believe that although the fields are
erroneous, the error distribution has to be such that errors become
more unlikely as their magnitude increases. For example, the family
name \textit{RODRIGEZ} is more likely to be a misrecording of
\textit{RODRIGUEZ} than of \textit{RAMIREZ}. Therefore, these truncation points
$\lambda_{fl}, l>0$, indicate that the probability of observing a
level of disagreement among coreferent pairs decreases as the
disagreement increases. For example, the probability $m_{\mathrm{Year},1}=\mathbb
{P}_1(\Gamma^{\mathrm{Year}}_{ij}=1|\Gamma^{\mathrm{Year}}_{ij}>0)$ is set to be minimum
0.9 a priori, that is, the probability that a coreferent pair disagrees
by one year (level one of disagreement, see Table~\ref{t:toyexample_compdata}) \emph{given} that it disagrees in year of
death (i.e., $\Gamma^{\mathrm{Year}}_{ij}>0$) should be at least 0.9. This
indicates that among all coreferent pairs that have disagreements in
year of death, we expect the majority to have the minimum disagreement,
which is one year ($\Gamma^{\mathrm{Year}}_{ij}=1$). Similarly, $m_{\mathrm{Year},2}$ is
set to be minimum 0.99 a priori, that is, the probability that a
coreferent pair disagrees by two or three years (level two of
disagreement, see Table~\ref{t:toyexample_compdata}) \emph{given} that
it disagrees by more than one year (i.e., $\Gamma^{\mathrm{Year}}_{ij}>1$)
should be at least 0.99. This prior specification constrains the prior
probability of the level three of disagreement (difference of four or
more years, see Table~\ref{t:toyexample_compdata}) to be very small
among coreferent pairs.

Finally, the prior for the field day of death has lower truncation
points since we believe this field to be more unreliable than the rest,
given that we do not expect witnesses to have been very accurate
reporting the exact date of the killings.

\subsection{Exploring the posterior sample of coreference partitions}

\begin{figure}

\includegraphics{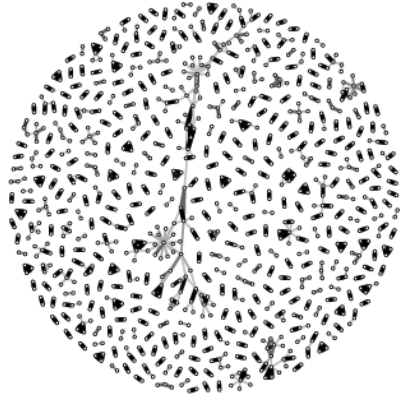}

\caption{The set $\mathcal{C}$ of candidate pairs for duplicate
detection. Each node represents a record, and two nodes appear
connected if their corresponding records are candidates to be
coreferent (i.e., not fixed as noncoreferent in the preprocessing
step). The color and width of the edges convey the same information:
The darker and thicker the edge, the larger the proportion of
partitions in the posterior sample that group the pair together.
Therefore, the lightest and thinnest edges indicate that those pairs
never appeared together, and the black and thickest edges indicate
those pairs were grouped together across all partitions in the
posterior sample.}\label{f:network_possible_links}
\end{figure}

%

We obtained a posterior sample of partitions of size 19,800 using the
Gibbs sampler and the implementation presented in the supplementary
material [\citet{SadinleAOASSupplement}]. For the sake of illustration,
in Figure~\ref{f:network_possible_links} we present a graph where each
node represents one record, and the existence of an edge indicates that
the pair was not fixed as noncoreferent in the preprocessing step, that
is, there is one edge per pair in $\mathcal{C}$. This graph was
obtained using the R package ``\texttt{igraph}'' [\citet{igraph}]. Our target
coreference partition can be thought of as a subgraph of this graph
composed by cliques. The sparsity of the graph in Figure~\ref{f:network_possible_links} illustrates the impact of fixing trivially
noncoreferent pairs in the preprocessing step: the number of pairs that
have to be resolved is small, and the possible set of partitions of the
file gets greatly constrained. In Figure~\ref{f:network_possible_links}
the color and the width of an edge are both proportional to the number
of times that the pair appears grouped together across the chain of
partitions. The thinnest and lightest edges indicate that the pair
never appeared together in the partitions of the chain, whereas the
thickest and black edges indicate that the pair appeared grouped
together in all the partitions of the chain. The black edges in Figure~\ref{f:network_possible_links} illustrate the property of the method of
ensuring transitive coreference decisions.

\begin{figure}

\includegraphics{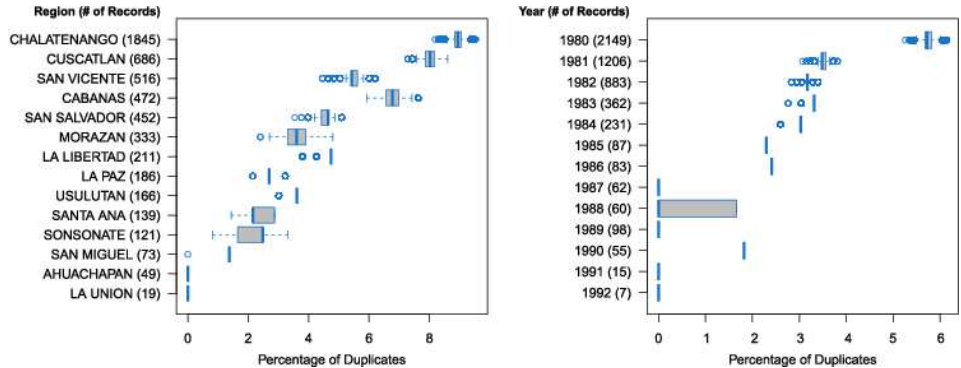}

\caption{Left panel: Percentage of duplicates per reported region of
death. The regions are ordered by the number of records. Right panel:
Percentage of duplicates per reported year of death, in chronological
order. In both cases the corresponding numbers of
records appear in parenthesis.}\label{f:resultsUNTRC}
\end{figure}

The output of our method is a posterior sample of possible coreference
partitions. Each of those partitions has a number of cells, which
represent unique entities, or, in this case, unique homicides. The
number of records minus the number of cells of a partition represents
the number of duplicates according to that partition. We can therefore
obtain a posterior distribution on this number. For the complete file,
which contained 5395 records, the posterior distribution on the number
of unique homicides has a mean and median of 5008, with a minimum of
4991, and a maximum of 5026 unique homicides, and a posterior 90\%
probability interval of $[5001, 5015]$, which corresponds to a
posterior interval on the percentage of duplicates of $[7.04, 7.30]$.
The rate of duplication greatly varies across different subsets of the
file. In Figure~\ref{f:resultsUNTRC} we summarize the posterior
distribution of the percentage of duplicates for subsets of the data
file corresponding to the different reported years and regions. The
left panel of Figure~\ref{f:resultsUNTRC} presents the regions of El
Salvador ordered by the number of records in the data file. We can
observe that the percentage of duplicates is correlated with the number
of homicides reported in that region: the more homicides reported, the
larger the proportion of duplicates. A similar relation can be observed
from the right panel of Figure~\ref{f:resultsUNTRC}, which shows the
percentage of duplicates per year.

%

\subsection{Evaluation of results and sensitivity analysis}

Although there is no ground truth for the UNTC data file, it is
important to have an idea of whether the results that we obtained are
reasonable at all. To this end, we took the UNTC records that reported
Cuscatl\'an and Ahuachap\'an as the regions of death (735 records), and
identified possible duplicates among them by hand. At this point, it is
important to clarify that we do not intend to treat these hand-labeled
records as ground truth, since they are also the product of our
subjective decisions, but rather we use them as a way to create a
sanity check for our results. The idea is to compare each partition in
the posterior sample with the hand-partitioned file subset in terms of
precision and recall.

We also would like to explore how sensitive our results are to small
changes in the prior truncation points that we chose. For this purpose,
we obtained two new posterior samples of partitions using two
alternative priors. We consider one prior more pessimistic and one more
optimistic than the one used in our application, in the sense that the
maximum amounts of error in the fields could be larger or smaller than
the ones implied by the prior truncation points set in Table~\ref{t:prior_ElSalvador}. These priors are obtained from subtracting/adding
0.02 to the prior truncation points of the $m_{fl}$ parameters in Table~\ref{t:prior_ElSalvador}, for $l=0,1$, and for all fields. For these
two additional priors we keep the same truncations of the $m_{f2}$ parameters.

\begin{figure}

\includegraphics{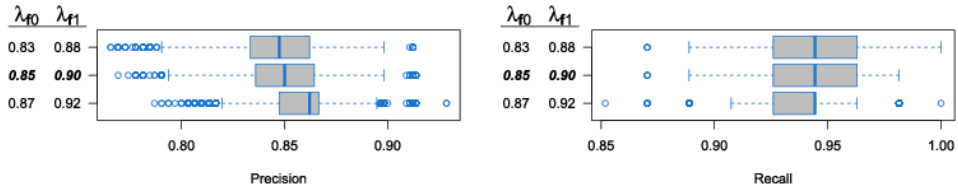}

\caption{Posterior distribution of precision (left panel) and recall
(right panel) computed with respect to hand-labeled records for two
regions of El Salvador. Results obtained under three different sets of
prior truncation points $\lambda_{fl}$ of the $m_{fl}$ parameters. The
$\lambda_{f2}$'s are fixed as in Table~\protect\ref{t:prior_ElSalvador}. The
prior truncation points used in the application to the UNTC data set
are indicated in bold italics.}
\label{f:sensitivity}
\end{figure}

In Figure~\ref{f:sensitivity} we summarize the posterior distributions
of precision and recall under the different priors considered here. We
can see that the precision of the method is somewhat sensitive to
changes in the prior truncation points and, although the recall is
somewhat robust, it starts to decay in the more optimistic scenario.
These results agree with the findings presented in Section~\ref{ss:simAustralia} for data files with a small number of fields. We
conclude that the prior employed in the application to the UNTC data
file achieves a good balance between precision and recall, since a more
optimistic prior would lead to less recall, and a more pessimistic
prior would lead to less precision.

Finally, we want to illustrate the issues that we would encounter if we
were using a model that outputs pairwise coreference decisions for the
UNTC data set. We implemented a two-components mixture model version of
the model presented in Section~\ref{s:modelindepcomps} to classify
the pairs in $\mathcal{C}$ into coreferent and noncoreferent pairs. The
mixture model is obtained by simply taking $\Delta_{ij}|p\stackrel
{\mathrm{i.i.d.}}{\sim} \operatorname{Bernoulli}(p)$, $i< j$, instead of treating $\bD$ as
the representation of a partition. We used Bayesian estimation of this
mixture model employing the same priors for the $m_{fl}$ and $u_{fl}$
parameters as in the application to the UNTC data set, and $p\sim\operatorname
{Uniform}(0,1)$. From running a Gibbs sampler for 100,000 iterations,
we obtained a posterior sample of $\Delta_{ij}$'s. The number of
nontransitive triplets varies between 69 and 564 across the Gibbs
iterations, which is not surprising given that this model treats the
$\Delta_{ij}$'s as independent. As we mentioned in the \hyperref[sec1]{Introduction} of
this article, if we wanted to use this mixture model approach, we would
have to implement some ad-hoc strategy to ensure transitivity of the
coreference decisions.

\section{Conclusions and future work}\label{s:conclusions}

We presented a novel, unsupervised approach to duplicate detection
problems. This approach improves over current methodology since it
guarantees transitive decisions, it allows us to incorporate prior
information on the amount of error in the fields, and it provides a
natural account for uncertainty of the coreference decisions in the
form of a posterior distribution. We showed that the method provides
reasonable results in an illustrative example and in a realistic
simulation study. The application of this methodology to detect
homicides reported multiple times to the Salvadoran UNTC indicates
that, with 90\% of probability, between 7.04\% and 7.30\% of those
reports are duplicates.

A number of improvements can be made to this methodology. For example,
the usage of field value frequencies would take into account that, for
instance, a~name that is relatively rare has more distinguishing power
than a common one [\citet{Winkler89Frequencies}]. Other extensions
include modeling dependencies between field comparisons, possibly
building on the work of \citet{LarsenRubin01}, and point estimation for
the coreference partition.

Our approach to duplicate detection is especially promising in the
context of multiple systems estimation of population sizes, which plays
an important role in human rights research [see \citet{LumPriceBanks13}]. It is important to note that the UNTC data file
does not cover all the deaths that occurred during the civil war of El
Salvador. Nevertheless, the combination of this source of information
with other data files on killings can provide a better account of the
lethal violence in El Salvador during the civil war. To pursue this
goal, our future work includes the extension of this methodology to
link multiple files, at the same time as finding duplicates within
them. Our Bayesian approach to this problem will allow us to
incorporate the uncertainty from record linkage and duplicate detection
into subsequent procedures, such as population size estimation.

\section*{Acknowledgments}
This document benefited from
discussions with the members of the CMU NCRN node. The author thanks
K.~Bokalders, S.~E. Fienberg, R.~C. Steorts, R.~B. Stern, the editor and
referees for helpful comments and suggestions that improved the quality
of this article; P.~Ball, A.~Hoover Green, J.~Klingner and M.~E. Price
from the Human Rights Data Analysis Group---HRDAG for providing the
UNTC data file and other important documentation; and P. Christen for
help with the generation of realistic synthetic data files.
%

\begin{supplement}
\stitle{Supplement to ``Detecting duplicates in a homicide registry
using\break a Bayesian partitioning approach''}
\slink[doi]{10.1214/14-AOAS779SUPP} 
\sdatatype{.pdf}
\sfilename{aoas779\_supp.pdf}
\sdescription{We provide a Gibbs sampler for the model presented in
Section~\ref{s:modelindepcomps}, a brief discussion on point
estimation of the coreference partition, we explain how we standardized
and compared Hispanic names and, finally, we present details on the
implementation of the Gibbs sampler for the application in
Section~\ref{s:UNTRCapplication}.}
\end{supplement}


%

\printaddresses

\begin{thebibliography}{45}

\bibitem[\protect\citeauthoryear{Bilenko et~al.}{2003}]{Bilenkoetal03}
\begin{barticle}[author]
\bauthor{\bsnm{Bilenko},~\bfnm{M.}\binits{M.}},
\bauthor{\bsnm{Mooney},~\bfnm{R.~J.}\binits{R.~J.}},
\bauthor{\bsnm{Cohen},~\bfnm{W.~W.}\binits{W.~W.}},
\bauthor{\bsnm{Ravikumar},~\bfnm{P.}\binits{P.}} \AND
\bauthor{\bsnm{Fienberg},~\bfnm{S.~E.}\binits{S.~E.}}
(\byear{2003}).
\btitle{{Adaptive name matching in information integration}}.
\bjournal{IEEE Intelligent Systems}
\bvolume{18}
\bpages{16--23}.
\end{barticle}
%
\bptok{imsref}%
\endbibitem

\bibitem[\protect\citeauthoryear{Buergenthal}{1994}]{Buergenthal94}
\begin{barticle}[author]
\bauthor{\bsnm{Buergenthal},~\bfnm{Thomas}\binits{T.}}
(\byear{1994}).
\btitle{{The United Nations Truth Commission for El Salvador}}.
\bjournal{Vanderbilt Journal of Transnational Law}
\bvolume{27}
\bpages{497--544}.
\end{barticle}
%
\bptok{imsref}%
\endbibitem

\bibitem[\protect\citeauthoryear{Buergenthal}{1996}]{Buergenthal96}
\begin{bincollection}[author]
\bauthor{\bsnm{Buergenthal},~\bfnm{Thomas}\binits{T.}}
(\byear{1996}).
\btitle{{La Comisi\'on de la Verdad para El Salvador}}.
In \bbooktitle{{Estudios Especializados de Derechos Humanos I}}
\bpages{11--62}.
\bpublisher{Instituto Interamericano de Derechos Humanos},
\blocation{San Jos{\'e}, Costa Rica}.
\end{bincollection}
%
\bptok{imsref}%
\endbibitem

\bibitem[\protect\citeauthoryear{Christen}{2005}]{Christen05}
\begin{binproceedings}[author]
\bauthor{\bsnm{Christen},~\bfnm{Peter}\binits{P.}}
(\byear{2005}).
\btitle{{Probabilistic data generation for deduplication and data linkage}}.
In \bbooktitle{Proceedings of the Sixth International Conference on Intelligent Data Engineering and Automated Learning (IDEAL'05)}
\bpages{109--116}.
\bpublisher{Springer},
\blocation{Berlin}.
\end{binproceedings}
%
\bptok{imsref}%
\endbibitem

\bibitem[\protect\citeauthoryear{Christen}{2012a}]{ChristenBook}
\begin{bbook}[author]
\bauthor{\bsnm{Christen},~\bfnm{Peter}\binits{P.}}
(\byear{2012}a).
\btitle{{Data Matching: Concepts and Techniques for Record Linkage, Entity Resolution, and Duplicate Detection}}.
\bpublisher{Springer},
\blocation{Berlin}.
\end{bbook}
%
\bptok{imsref}%
\endbibitem

\bibitem[\protect\citeauthoryear{Christen}{2012b}]{Christen12}
\begin{barticle}[author]
\bauthor{\bsnm{Christen},~\bfnm{Peter}\binits{P.}}
(\byear{2012}b).
\btitle{{A survey of indexing techniques for scalable record linkage and deduplication}}.
\bjournal{IEEE Transactions on Knowledge and Data Engineering}
\bvolume{24}
\bpages{1537--1555}.
\end{barticle}
\bptok{imsref}%
\endbibitem

\bibitem[\protect\citeauthoryear{Christen and Pudjijono}{2009}]{ChristenPudjijono09}
\begin{bincollection}[author]
\bauthor{\bsnm{Christen},~\bfnm{Peter}\binits{P.}} \AND
\bauthor{\bsnm{Pudjijono},~\bfnm{Agus}\binits{A.}}
(\byear{2009}).
\btitle{{Accurate synthetic generation of realistic personal information}}.
In \bbooktitle{Advances in Knowledge Discovery and Data Mining}
(\beditor{\bfnm{Thanaruk}\binits{T.}~\bsnm{Theeramunkong}},
\beditor{\bfnm{Boonserm}\binits{B.}~\bsnm{Kijsirikul}},
\beditor{\bfnm{Nick}\binits{N.}~\bsnm{Cercone}} \AND
\beditor{\bfnm{Tu-Bao}\binits{T.-B.}~\bsnm{Ho}}, eds.).
\bseries{Lecture Notes in Computer Science}
\bvolume{5476}
\bpages{507--514}.
\bpublisher{Springer},
\blocation{Berlin}.
\end{bincollection}
%
\bptok{imsref}%
\endbibitem

\bibitem[\protect\citeauthoryear{Christen and Vatsalan}{2013}]{ChristenVatsalan13}
\begin{binproceedings}[author]
\bauthor{\bsnm{Christen},~\bfnm{Peter}\binits{P.}} \AND
\bauthor{\bsnm{Vatsalan},~\bfnm{Dinusha}\binits{D.}}
(\byear{2013}).
\btitle{{Flexible and extensible generation and corruption of personal data}}.
In \bbooktitle{Proceedings of the ACM International Conference on Information
and Knowledge Management (CIKM 2013)}.
\bpublisher{ACM},
\blocation{New York}.
\end{binproceedings}
\bptok{imsref}%
\endbibitem

\bibitem[\protect\citeauthoryear{{Commission on the Truth for El Salvador}}{1993}]{CTElSalvador93}
\begin{bmisc}[author]
\borganization{Commission on the Truth for El Salvador}
(\byear{1993}).
\bhowpublished{From madness to hope: The 12-year war in El Salvador:
Report of the Commission on the Truth for El Salvador.
Available at \url{http://www.usip.org/files/file/ElSalvador-Report.pdf}
[Accessed October 15, 2014].
UN Security Council}.
\end{bmisc}
%
\bptok{imsref}%
\endbibitem

\bibitem[\protect\citeauthoryear{Csardi and Nepusz}{2006}]{igraph}
\begin{barticle}[author]
\bauthor{\bsnm{Csardi},~\bfnm{Gabor}\binits{G.}} \AND
\bauthor{\bsnm{Nepusz},~\bfnm{Tamas}\binits{T.}}
(\byear{2006}).
\btitle{{The igraph software package for complex network research}}.
\bjournal{InterJournal}
\bvolume{Complex Systems}
\bpages{1695}.
\end{barticle}
\bptok{imsref}%
\endbibitem

\bibitem[\protect\citeauthoryear{Elmagarmid, Ipeirotis and Verykios}{2007}]{Elmagarmidetal07}
\begin{barticle}[author]
\bauthor{\bsnm{Elmagarmid},~\bfnm{Ahmed~K.}\binits{A.~K.}},
\bauthor{\bsnm{Ipeirotis},~\bfnm{Panagiotis~G.}\binits{P.~G.}} \AND
\bauthor{\bsnm{Verykios},~\bfnm{Vassilios~S.}\binits{V.~S.}}
(\byear{2007}).
\btitle{{Duplicate record detection: A survey}}.
\bjournal{IEEE Transactions on Knowledge and Data Engineering}
\bvolume{19}
\bpages{1--16}.
\end{barticle}
%
\bptok{imsref}%
\endbibitem

\bibitem[\protect\citeauthoryear{Fay}{2004}]{Fay04}
\begin{binproceedings}[author]
\bauthor{\bsnm{Fay},~\bfnm{Robert~E.}\binits{R.~E.}}
(\byear{2004}).
\btitle{{An analysis of person duplication in census 2000}}.
In \bbooktitle{Proceedings of the Section on Survey Research Methods}
\bpages{3478--3485}.
\bpublisher{Amer. Statist. Assoc.},
\blocation{Alexandria, VA}.
\end{binproceedings}
%
\bptok{imsref}%
\endbibitem

\bibitem[\protect\citeauthoryear{Fellegi and Sunter}{1969}]{FellegiSunter69}
\begin{barticle}[author]
\bauthor{\bsnm{Fellegi},~\bfnm{Ivan~P.}\binits{I.~P.}} \AND
\bauthor{\bsnm{Sunter},~\bfnm{Alan~B.}\binits{A.~B.}}
(\byear{1969}).
\btitle{{A theory for record linkage}}.
\bjournal{J. Amer. Statist. Assoc.}
\bvolume{64}
\bpages{1183--1210}.
\end{barticle}
%
\bptok{imsref}%
\endbibitem

\bibitem[\protect\citeauthoryear{Fern{\'a}ndez and Garc{\'i}a}{2003}]{FernandezGarcia03}
\begin{barticle}[author]
\bauthor{\bsnm{Fern{\'a}ndez},~\bfnm{Esteve}\binits{E.}} \AND
\bauthor{\bsnm{Garc{\'i}a},~\bfnm{Ana~M.}\binits{A.~M.}}
(\byear{2003}).
\btitle{{Accuracy of referencing of Spanish names in Medline}}.
\bjournal{The Lancet}
\bvolume{361}
\bpages{351--352}.
\end{barticle}
\bptok{imsref}%
\endbibitem

\bibitem[\protect\citeauthoryear{Fortini et~al.}{2001}]{Fortinietal01}
\begin{barticle}[author]
\bauthor{\bsnm{Fortini},~\bfnm{M.}\binits{M.}},
\bauthor{\bsnm{Liseo},~\bfnm{B.}\binits{B.}},
\bauthor{\bsnm{Nuccitelli},~\bfnm{A.}\binits{A.}} \AND
\bauthor{\bsnm{Scanu},~\bfnm{M.}\binits{M.}}
(\byear{2001}).
\btitle{{On Bayesian record linkage}}.
\bjournal{Researh in Official Statistics}
\bvolume{4}
\bpages{185--198}.
\end{barticle}
%
\bptok{imsref}%
\endbibitem

\bibitem[\protect\citeauthoryear{Fortini et~al.}{2002}]{Fortinietal02}
\begin{binproceedings}[author]
\bauthor{\bsnm{Fortini},~\bfnm{M.}\binits{M.}},
\bauthor{\bsnm{Nuccitelli},~\bfnm{A.}\binits{A.}},
\bauthor{\bsnm{Liseo},~\bfnm{B.}\binits{B.}} \AND
\bauthor{\bsnm{Scanu},~\bfnm{M.}\binits{M.}}
(\byear{2002}).
\btitle{{Modeling issues in record linkage: A Bayesian perspective}}.
In \bbooktitle{Proceedings of the Section on Survey Research Methods}
\bpages{1008--1013}.
\bpublisher{Amer. Statist. Assoc.},
\blocation{Alexandria, VA}.
\end{binproceedings}
%
\bptok{imsref}%
\endbibitem

\bibitem[\protect\citeauthoryear{Gutman, Afendulis and Zaslavsky}{2013}]{Gutmanetal13}
\begin{barticle}[mr]
\bauthor{\bsnm{Gutman},~\bfnm{Roee}\binits{R.}},
\bauthor{\bsnm{Afendulis},~\bfnm{Christopher~C.}\binits{C.~C.}} \AND
\bauthor{\bsnm{Zaslavsky},~\bfnm{Alan~M.}\binits{A.~M.}}
(\byear{2013}).
\btitle{A {B}ayesian procedure for file linking to analyze end-of-life medical costs}.
\bjournal{J. Amer. Statist. Assoc.}
\bvolume{108}
\bpages{34--47}.
\bid{doi={10.1080/01621459.2012.726889}, issn={0162-1459}, mr={3174601}}
\end{barticle}
\bptok{imsref}%
\endbibitem

\bibitem[\protect\citeauthoryear{Herzog, Scheuren and Winkler}{2007}]{HerzogScheurenWinkler07}
\begin{bbook}[author]
\bauthor{\bsnm{Herzog},~\bfnm{Thomas~N.}\binits{T.~N.}},
\bauthor{\bsnm{Scheuren},~\bfnm{Fritz~J.}\binits{F.~J.}} \AND
\bauthor{\bsnm{Winkler},~\bfnm{William~E.}\binits{W.~E.}}
(\byear{2007}).
\btitle{{Data Quality and Record Linkage Techniques}}.
\bpublisher{Springer},
\blocation{New York}.
\end{bbook}
\bptok{imsref}%
\endbibitem

\bibitem[\protect\citeauthoryear{{Hoover Green}}{2011}]{Amelia11}
\begin{bmisc}[author]
\bauthor{\bsnm{{Hoover Green}},~\bfnm{Amelia}\binits{A.}}
(\byear{2011}).
\bhowpublished{Repertoires of violence against noncombatants:
The role of armed group institutions and ideologies.
Ph.D. thesis, Yale Univ}.
\end{bmisc}
%
\bptok{imsref}%
\endbibitem

\bibitem[\protect\citeauthoryear{Hsu et~al.}{2000}]{Hsuetal00}
\begin{binproceedings}[author]
\bauthor{\bsnm{Hsu},~\bfnm{Wynne}\binits{W.}},
\bauthor{\bsnm{Lee},~\bfnm{Mong~Li}\binits{M.~L.}},
\bauthor{\bsnm{Liu},~\bfnm{Bing}\binits{B.}} \AND
\bauthor{\bsnm{Ling},~\bfnm{Tok~Wang}\binits{T.~W.}}
(\byear{2000}).
\btitle{{Exploration mining in diabetic patients databases: Findings
and conclusions}}.
In \bbooktitle{Proceedings of the Sixth ACM SIGKDD International Conference on Knowledge Discovery and Data Mining (KDD '00)}
\bpages{430--436}.
\bpublisher{ACM},
\blocation{New York}.
\end{binproceedings}
%
\bptok{imsref}%
\endbibitem

\bibitem[\protect\citeauthoryear{Jaro}{1989}]{Jaro89}
\begin{barticle}[author]
\bauthor{\bsnm{Jaro},~\bfnm{Matthew~A.}\binits{M.~A.}}
(\byear{1989}).
\btitle{{Advances in record-linkage methodology as applied to matching the 1985 census of Tampa, Florida}}.
\bjournal{J. Amer. Statist. Assoc.}
\bvolume{84}
\bpages{414--420}.
\end{barticle}
%
\bptok{imsref}%
\endbibitem

\bibitem[\protect\citeauthoryear{Keener, Rothman and Starr}{1987}]{Keeneretal87}
\begin{barticle}[mr]
\bauthor{\bsnm{Keener},~\bfnm{Robert}\binits{R.}},
\bauthor{\bsnm{Rothman},~\bfnm{Edward}\binits{E.}} \AND
\bauthor{\bsnm{Starr},~\bfnm{Norman}\binits{N.}}
(\byear{1987}).
\btitle{Distributions on partitions}.
\bjournal{Ann. Statist.}
\bvolume{15}
\bpages{1466--1481}.
\bid{doi={10.1214/aos/1176350604}, issn={0090-5364}, mr={0913568}}
\end{barticle}
%
\bptok{imsref}%
\endbibitem

\bibitem[\protect\citeauthoryear{Larsen}{2002}]{Larsen02}
\begin{binproceedings}[author]
\bauthor{\bsnm{Larsen},~\bfnm{Michael~D.}\binits{M.~D.}}
(\byear{2002}).
\btitle{{Comments on hierarchical Bayesian record linkage}}.
In \bbooktitle{Proceedings of the Section on Survey Research Methods}
\bpages{1995--2000}.
\bpublisher{Amer. Statist. Assoc.},
\blocation{Alexandria, VA}.
\end{binproceedings}
\bptok{imsref}%
\endbibitem

\bibitem[\protect\citeauthoryear{Larsen}{2005}]{Larsen05}
\begin{binproceedings}[author]
\bauthor{\bsnm{Larsen},~\bfnm{Michael~D.}\binits{M.~D.}}
(\byear{2005}).
\btitle{{Advances in record linkage theory: Hierarchical Bayesian record linkage theory}}.
In \bbooktitle{Proceedings of the Section on Survey Research Methods}
\bpages{3277--3284}.
\bpublisher{Amer. Statist. Assoc.},
\blocation{Alexandria, VA}.
\end{binproceedings}
%
\bptok{imsref}%
\endbibitem

\bibitem[\protect\citeauthoryear{Larsen}{2012}]{Larsen12}
\begin{bmisc}[author]
\bauthor{\bsnm{Larsen},~\bfnm{Michael~D.}\binits{M.~D.}}
(\byear{2012}).
\bhowpublished{An experiment with hierarchical Bayesian record linkage.
Preprint. Available at \url{http://arxiv.org/abs/1212.5203}.}
\end{bmisc}
%
\bptok{imsref}%
\endbibitem

\bibitem[\protect\citeauthoryear{Larsen and Rubin}{2001}]{LarsenRubin01}
\begin{barticle}[mr]
\bauthor{\bsnm{Larsen},~\bfnm{Michael~D.}\binits{M.~D.}} \AND
\bauthor{\bsnm{Rubin},~\bfnm{Donald~B.}\binits{D.~B.}}
(\byear{2001}).
\btitle{Iterative automated record linkage using mixture models}.
\bjournal{J. Amer. Statist. Assoc.}
\bvolume{96}
\bpages{32--41}.
\bid{doi={10.1198/016214501750332956}, issn={0162-1459}, mr={1973781}}
\end{barticle}
\bptok{imsref}%
\endbibitem

\bibitem[\protect\citeauthoryear{Little and Rubin}{2002}]{LittleRubin02}
\begin{bbook}[mr]
\bauthor{\bsnm{Little},~\bfnm{Roderick~J.~A.}\binits{R.~J.~A.}} \AND
\bauthor{\bsnm{Rubin},~\bfnm{Donald~B.}\binits{D.~B.}}
(\byear{2002}).
\btitle{Statistical Analysis with Missing Data},
\bedition{2nd} ed.
\bpublisher{Wiley},
\blocation{Hoboken, NJ}.
\bid{mr={1925014}}
\end{bbook}
%
\bptok{imsref}%
\endbibitem

\bibitem[\protect\citeauthoryear{Lum, Price and Banks}{2013}]{LumPriceBanks13}
\begin{barticle}[author]
\bauthor{\bsnm{Lum},~\bfnm{Kristian}\binits{K.}},
\bauthor{\bsnm{Price},~\bfnm{Megan~Emily}\binits{M.~E.}} \AND
\bauthor{\bsnm{Banks},~\bfnm{David}\binits{D.}}
(\byear{2013}).
\btitle{{Applications of multiple systems estimation in human rights research}}.
\bjournal{Amer. Statist.}
\bvolume{67}
\bpages{191--200}.
\end{barticle}
%
\bptok{imsref}%
\endbibitem

\bibitem[\protect\citeauthoryear{Marshall}{2008}]{Marshall08}
\begin{binproceedings}[author]
\bauthor{\bsnm{Marshall},~\bfnm{Leah}\binits{L.}}
(\byear{2008}).
\btitle{{Potential duplicates in the census: Methodology and selection of cases for followup}}.
In \bbooktitle{Proceedings of the Section on Survey Research Methods}
\bpages{4237--4244}.
\bpublisher{Amer. Statist. Assoc.},
\blocation{Alexandria, VA}.
\end{binproceedings}
%
\bptok{imsref}%
\endbibitem

\bibitem[\protect\citeauthoryear{Matsakis}{2010}]{Matsakis10}
\begin{bmisc}[author]
\bauthor{\bsnm{Matsakis},~\bfnm{Nicholas~Elias}\binits{N.~E.}}
(\byear{2010}).
\bhowpublished{Active duplicate detection with Bayesian nonparametric models.
Ph.D. thesis, Massachusetts Institute of Technology}.
\end{bmisc}
%
\bptok{imsref}%
\endbibitem

\bibitem[\protect\citeauthoryear{McCullagh}{2011}]{McCullagh11}
\begin{binbook}[author]
\bauthor{\bsnm{McCullagh},~\bfnm{Peter}\binits{P.}}
(\byear{2011}).
\btitle{{Random permutations and partition models}}.
In \bbooktitle{{International Encyclopedia of Statistical Science}}
\bpages{1170--1177}.
\bpublisher{Springer},
\blocation{Berlin}.
\end{binbook}
%
\bptok{imsref}%
\endbibitem

\bibitem[\protect\citeauthoryear{Miller, Frawley and Sayward}{2000}]{Milleretal00}
\begin{barticle}[author]
\bauthor{\bsnm{Miller},~\bfnm{Perry~L.}\binits{P.~L.}},
\bauthor{\bsnm{Frawley},~\bfnm{Sandra~J.}\binits{S.~J.}} \AND
\bauthor{\bsnm{Sayward},~\bfnm{Frederick~G.}\binits{F.~G.}}
(\byear{2000}).
\btitle{{IMM/Scrub: A domain-specific tool for the deduplication of vaccination history records in childhood immunization registries}}.
\bjournal{Computers and Biomedical Research}
\bvolume{33}
\bpages{126--143}.
\end{barticle}
%
\bptok{imsref}%
\endbibitem

\bibitem[\protect\citeauthoryear{{R Core Team}}{2013}]{R13}
\begin{bmisc}[author]
\borganization{R Core Team}
(\byear{2013}).
\bhowpublished{\textit{R: A Language and Environment for Statistical Computing}.
R Foundation for Statistical Computing, Vienna, Austria}.
\end{bmisc}
\bptok{imsref}%
\endbibitem

\bibitem[\protect\citeauthoryear{Rota}{1964}]{Rota64}
\begin{barticle}[mr]
\bauthor{\bsnm{Rota},~\bfnm{Gian-Carlo}\binits{G.-C.}}
(\byear{1964}).
\btitle{The number of partitions of a set}.
\bjournal{Amer. Math. Monthly}
\bvolume{71}
\bpages{498--504}.
\bid{issn={0002-9890}, mr={0161805}}
\end{barticle}
%
\bptok{imsref}%
\endbibitem

\bibitem[\protect\citeauthoryear{Ruiz-P{\'e}rez, L{\'o}pez-C{\'o}zar and Jim{\'e}nez-Contreras}{2002}]{RuizPerezetal02}
\begin{barticle}[author]
\bauthor{\bsnm{Ruiz-P{\'e}rez},~\bfnm{R.}\binits{R.}},
\bauthor{\bsnm{L{\'o}pez-C{\'o}zar},~\bfnm{E.~Delgado}\binits{E.~D.}} \AND
\bauthor{\bsnm{Jim{\'e}nez-Contreras},~\bfnm{E.}\binits{E.}}
(\byear{2002}).
\btitle{{Spanish personal name variations in national and international biomedical databases: Implications for information retrieval and bibliometric studies}}.
\bjournal{Journal of the Medical Library Association}
\bvolume{90}
\bpages{411--430}.
\end{barticle}
%
\bptok{imsref}%
\endbibitem

\bibitem[\protect\citeauthoryear{Sadinle}{2014}]{SadinleAOASSupplement}
\begin{bmisc}[author]
{\bauthor{\bsnm{Sadinle},~\binits{M.}}}
(\byear{2014}).
\bhowpublished{Supplement to ``Detecting duplicates in a homicide
registry using a Bayesian
partitioning approach.''
DOI:\doiurl{10.1214/14-AOAS779SUPP}}.
\bptok{imsref}%
\end{bmisc}
\endbibitem

\bibitem[\protect\citeauthoryear{Sadinle and Fienberg}{2013}]{SadinleFienberg13}
\begin{barticle}[mr]
\bauthor{\bsnm{Sadinle},~\bfnm{Mauricio}\binits{M.}} \AND
\bauthor{\bsnm{Fienberg},~\bfnm{Stephen~E.}\binits{S.~E.}}
(\byear{2013}).
\btitle{A generalized {F}ellegi--{S}unter framework for multiple record linkage with application to homicide record systems}.
\bjournal{J. Amer. Statist. Assoc.}
\bvolume{108}
\bpages{385--397}.
\bid{doi={10.1080/01621459.2012.757231}, issn={0162-1459}, mr={3174628}}
\end{barticle}
\bptok{imsref}%
\endbibitem

\bibitem[\protect\citeauthoryear{Sariyar and Borg}{2010}]{SariyarBorg10}
\begin{barticle}[author]
\bauthor{\bsnm{Sariyar},~\bfnm{Murat}\binits{M.}} \AND
\bauthor{\bsnm{Borg},~\bfnm{Andreas}\binits{A.}}
(\byear{2010}).
\btitle{{The RecordLinkage package: Detecting errors in data}}.
\bjournal{The R Journal}
\bvolume{2}
\bpages{61--67}.
\end{barticle}
\bptok{imsref}%
\endbibitem

\bibitem[\protect\citeauthoryear{Sariyar, Borg and Pommerening}{2009}]{Sariyaretal09}
\begin{barticle}[pbm]
\bauthor{\bsnm{Sariyar},~\bfnm{Murat}\binits{M.}},
\bauthor{\bsnm{Borg},~\bfnm{A.}\binits{A.}} \AND
\bauthor{\bsnm{Pommerening},~\bfnm{K.}\binits{K.}}
(\byear{2009}).
\btitle{Evaluation of record linkage methods for iterative insertions}.
\bjournal{Methods Inf. Med.}
\bvolume{48}
\bpages{429--437}.
\bid{doi={10.3414/ME9238}, issn={0026-1270}, pii={9238}, pmid={19696952}}
\end{barticle}
%
\bptok{imsref}%
\endbibitem

\bibitem[\protect\citeauthoryear{Sariyar, Borg and Pommerening}{2012}]{Sariyaretal12MissingValues}
\begin{barticle}[author]
\bauthor{\bsnm{Sariyar},~\bfnm{M.}\binits{M.}},
\bauthor{\bsnm{Borg},~\bfnm{A.}\binits{A.}} \AND
\bauthor{\bsnm{Pommerening},~\bfnm{K.}\binits{K.}}
(\byear{2012}).
\btitle{{Missing values in deduplication of electronic patient data}}.
\bjournal{Journal of the American Medical Informatics Association}
\bvolume{19}
\bpages{e76--e82}.
\end{barticle}
%
\bptok{imsref}%
\endbibitem

\bibitem[\protect\citeauthoryear{Steorts, Hall and Fienberg}{2013}]{Steortsetal13}
\begin{bmisc}[author]
\bauthor{\bsnm{Steorts},~\bfnm{Rebecca~C.}\binits{R.~C.}},
\bauthor{\bsnm{Hall},~\bfnm{Rob}\binits{R.}} \AND
\bauthor{\bsnm{Fienberg},~\bfnm{Stephen~E.}\binits{S.~E.}}
(\byear{2013}).
\bhowpublished{A Bayesian approach to graphical record linkage and
deduplication.
Preprint. Available at \url{http://arxiv.org/abs/1312.4645}.}
\end{bmisc}
\bptok{imsref}%
\endbibitem

\bibitem[\protect\citeauthoryear{Tancredi and Liseo}{2011}]{TancrediLiseo11}
\begin{barticle}[mr]
\bauthor{\bsnm{Tancredi},~\bfnm{Andrea}\binits{A.}} \AND
\bauthor{\bsnm{Liseo},~\bfnm{Brunero}\binits{B.}}
(\byear{2011}).
\btitle{A hierarchical {B}ayesian approach to record linkage and population size problems}.
\bjournal{Ann. Appl. Stat.}
\bvolume{5}
\bpages{1553--1585}.
\bid{doi={10.1214/10-AOAS447}, issn={1932-6157}, mr={2849786}}
\end{barticle}
%
\bptok{imsref}%
\endbibitem

\bibitem[\protect\citeauthoryear{Winkler}{1988}]{Winkler88}
\begin{binproceedings}[author]
\bauthor{\bsnm{Winkler},~\bfnm{W.~E.}\binits{W.~E.}}
(\byear{1988}).
\btitle{{Using the EM algorithm for weight computation in the Fellegi--Sunter
model of record linkage}}.
In \bbooktitle{Proceedings of the Section on Survey Research Methods}
\bpages{667--671}.
\bpublisher{Amer. Statist. Assoc.},
\blocation{Alexandria, VA}.
\end{binproceedings}
%
\bptok{imsref}%
\endbibitem

\bibitem[\protect\citeauthoryear{Winkler}{1989}]{Winkler89Frequencies}
\begin{binproceedings}[author]
\bauthor{\bsnm{Winkler},~\bfnm{W.~E.}\binits{W.~E.}}
(\byear{1989}).
\btitle{{Frequency-based matching in the Fellegi--Sunter model of record linkage}}.
In \bbooktitle{Proceedings of the Section on Survey Research Methods}
\bpages{778--783}.
\bpublisher{Amer. Statist. Assoc.},
\blocation{Alexandria, VA}.
\end{binproceedings}
%
\bptok{imsref}%
\endbibitem

\bibitem[\protect\citeauthoryear{Winkler}{1990}]{Winkler90Strings}
\begin{binproceedings}[author]
\bauthor{\bsnm{Winkler},~\bfnm{W.~E.}\binits{W.~E.}}
(\byear{1990}).
\btitle{{String comparator metrics and enhanced decision rules in the
Fellegi--Sunter model of record linkage}}.
In \bbooktitle{Proceedings of the Section on Survey Research Methods}
\bpages{354--359}.
\bpublisher{Amer. Statist. Assoc.},
\blocation{Alexandria, VA}.
\end{binproceedings}
\bptok{imsref}%
\endbibitem
\end{thebibliography}
\end{document}